\documentclass[a4paper,11pt]{article}
\usepackage{jcappub}

\pdfoutput=1

\usepackage{tabularx,booktabs}
\newcolumntype{Y}{>{\centering\arraybackslash}X}

\usepackage{fontawesome}
\usepackage[utf8]{inputenc}
\usepackage[font=small,labelfont=bf,tableposition=top]{caption}
\usepackage{float}
\usepackage{adjustbox}
\usepackage{arydshln}
\usepackage{empheq}
\newcommand\exactfbox[1]{\fbox{\hskip1em#1\hskip1em}}

\usepackage{amsmath,amssymb,amsbsy,amstext, amsthm, simplewick, amsfonts, bm}
\usepackage{hyperref}
\usepackage[normalem]{ulem}
\usepackage{graphicx}
\usepackage{siunitx}
\usepackage{upgreek}
\usepackage{framed}
\usepackage{wrapfig}
\usepackage{multirow}
\usepackage{subfig}
\usepackage{bbm}
\usepackage[nameinlink, capitalise]{cleveref}
\usepackage[svgnames,dvipsnames,x11names]{xcolor}
\usepackage{wrapfig}
\usepackage{soul}
\usepackage[htt]{hyphenat}
\allowdisplaybreaks

\usepackage[top=2.5cm,bottom=3.cm,right=2.5cm,left=2.5cm]{geometry}
\def\d{\partial}

\def\dd{{\rm d}}
\def\ln{{\rm ln}}

\def\be{\begin{equation}}
\def\ee{\end{equation}}
\def\bea{\begin{eqnarray}}
\def\eea{\end{eqnarray}}
\def\ba{\begin{align}}

\def\bi{\begin{itemize}}
\def\ei{\end{itemize}}

\def\hn{{\hat n}}
\def\bu{{\bf u}}
\def\bv{{\bf v}}
\def\bs{{\bf s}}
\def\br{{\bf r}}
\def\cH{{\mathcal H}}

\def\bx{{\bf x}}
\def\bk{{\bf k}}
\def\bq{{\bf q}}

\def\kmq{{|\bk-\bq|}}

\def\cG{{\mathcal G}}
\def\cF{{\mathcal F}}

\def\cI{{\mathcal I}}
\def\cJ{{\mathcal J}}
\def\cN{{\mathcal N}}

\title{\fontsize{20}{32}\selectfont{CLASS-OneLoop: Accurate and Unbiased \\ Inference from Spectroscopic Galaxy Surveys \vspace{0.1in}}} 

\author[a]{\fontsize{13.84}{25}\selectfont Dennis Linde}
\author[b,c]{, Azadeh Moradinezhad Dizgah}
\author[d]{, Christian Radermacher}
\author[d]{, Santiago Casas}
\author[d]{, Julien Lesgourgues  \vspace{0.1in}}

\affiliation[a]{Department of Mathematical, Physical and Computer Sciences, University of Parma, 43124 Parma, Italy}
\affiliation[b]{Laboratoire d’Annecy de Physique Théorique (CNRS/USMB), F-74940 Annecy, France}
\affiliation[c]{D\'epartement de Physique Th\'eorique, Universit\'e de Gen\`eve, 24 quai Ernest Ansermet, \\ 1211 Gen\`eva 4, Switzerland}
\affiliation[d]{Institute for Theoretical Particle Physics and Cosmology (TTK), RWTH Aachen University,
D-52056 Aachen, Germany}

\emailAdd{dennis.linde@unipr.it}
\emailAdd{azadeh.moradinezhad@lapth.cnrs.fr}
\emailAdd{christian.radermacher1@rwth-aachen.de}
\emailAdd{casas@physik.rwth-aachen.de}
\emailAdd{lesgourg@physik.rwth-aachen.de}

\abstract{The power spectrum is the most commonly applied summary statistics to extract cosmological information from the observed three-dimensional distribution of galaxies in spectroscopic surveys. We present {\sc CLASS-OneLoop}, a new numerical tool, fully integrated into the Boltzmann code {\sc CLASS}, enabling the calculation of the one-loop power spectrum of biased tracers in spectroscopic surveys. Built upon the Eulerian moment expansion framework for redshift-space distortions, the implemented model incorporates a complete set of nonlinear biases, counterterms, and stochastic contributions, and includes the infrared resummation and the Alcock-Paczynski effect. The code features an evaluation of the loops by either direct numerical integration or Fast Fourier Transform, and employs a fast-slow parameter decomposition, which is essential for accelerating MCMC runs. After presenting performance and validation tests, as an illustration of the capabilities of the code, we apply it to fit the measured redshift-space halo power spectrum wedges on a $\Lambda$CDM subset of the {\sc AbacusSummit} simulation suite and considering scales up to $k_{\rm max} = 0.3\,h/$Mpc. We find that the one-loop model adeptly recovers the fiducial cosmology of the simulation, while a simplified model commonly used in the literature for sensitivity forecasts yields significantly biased results. Furthermore, we conduct Monte Carlo Markov Chain (MCMC) forecasts for a DESI-like survey, considering a model with a dynamical dark energy component. Our results demonstrate the ability to independently constrain cosmological and nuisance parameters, even in the presence of a large parameter space with twenty-nine variables.}

\begin{document}

\begin{flushright}
TTK-24-08
\end{flushright}

\maketitle

\newpage
\section{Introduction}

The cosmic large-scale structure (LSS) is formed under the influence of gravity from quantum fluctuations generated in the primordial Universe. As such, the LSS contains invaluable information about the initial conditions of the Universe, its constituents, and its evolution. Building upon the spectacular success of latest redshift galaxy surveys, in particular Sloan Digital Sky Surveys (BOSS \& eBOSS) \href{https://www.sdss4.org/surveys/}{\faGlobe},\footnote{\url{https://www.sdss4.org/surveys/}} in the coming years, the stage-IV wide-field spectroscopic surveys, namely the Dark Energy Spectroscopic Instrument (DESI) \cite{Desjacques:2016bnm}, {\it Euclid} \cite{Amendola:2016saw}, Nancy Grace Roman Space Telescope \cite{Wang:2021oec}, and SPHEREx \cite{SPHEREx:2014bgr} will provide highly detailed three-dimensional maps of a large volume of the Universe. These datasets promise to deepen our understanding of early- and late-time Universe by placing stringent constraints on the $\Lambda$CDM model and potentially offering discovery space for new physics.

Extracting cosmological information from the upcoming data with unprecedented precision imposes stringent requirements on the accuracy of the theoretical model describing the observables to match the precision of the data. Additionally, efficient numerical algorithms are essential for computing the theoretical model and conducting likelihood analysis. Until very recently, due to limitations in these areas, standard analyses of large-scale structure (LSS) data focused primarily on extracting cosmological insights through Baryon Acoustic Oscillations (BAO) and redshift-space distortions (RSD) measurements, obtained by compressing the information contained in 2-point statistics (e.g.,\cite{BOSS:2016wmc,eBOSS:2020yzd}). 

For biased tracers of dark matter, such as galaxies and quasars, modeling three sources of non-linearities presents a significant challenge: one needs to account for the gravitational evolution of dark matter distribution, the biasing of tracers relative to the dark matter distribution, and the influence of the peculiar velocities of tracers. Recent theoretical advancements have resulted in a consistent model for the clustering statistics of dark matter and its biased tracers within perturbation theory \cite{McDonald:2006mx, Baumann:2010tm, Carrasco:2012cv, Senatore:2014eva,Senatore:2014via,Senatore:2014vja, Vlah:2015sea, Blas:2015qsi, Ivanov:2018gjr, Eggemeier:2018qae, Cabass:2022avo}. Simultaneously, the development of efficient algorithms for computing model predictions \cite{McEwen:2016fjn,Schmittfull:2016jsw,Simonovic:2017mhp} has enabled analyses of existing data (from BOSS and eBOSS surveys) to constrain the $\Lambda$CDM model and some of its extensions using the full shape of the redshift-space galaxy power spectrum \cite{Ivanov:2019hqk, DAmico:2019fhj, Philcox:2020vvt, DAmico:2020kxu, Chen:2021wdi,Moretti:2023drg} and bispectrum \cite{Philcox:2021kcw, Philcox:2022frc, Cabass:2022wjy, Cabass:2022ymb, DAmico:2022osl, DAmico:2022gki}. These analyses have yielded robust cosmological constraints, largely independent of the CMB data, in contrast to conventional analyses of galaxy clustering data based on BAO and RSD measurements, which assume the overall shape of the matter power spectrum to be fixed by CMB measurements (see e.g., \cite{Brieden:2021edu, Maus:2023rtr} for a discussion of the differences between the two approaches). As such, these new analyses allow to test the consistency of the constraints inferred from data on the high- and low-redshift Universe. 

Thanks to these recent developments, several software packages are now publicly available and have been used to analyze synthetic or real galaxy data. The list includes {\sc CLASS-PT} \cite{Chudaykin:2020aoj} \href{https://github.com/Michalychforever/CLASS-PT/}{\faGithub},\footnote{\url{https://github.com/Michalychforever/CLASS-PT/}} PyBird \cite{DAmico:2020kxu} \href{https://github.com/pierrexyz/pybird/}{\faGithub},\footnote{\url{https://github.com/pierrexyz/pybird/}} and Velocileptor \cite{Chen:2020fxs,Chen:2020zjt} \href{https://github.com/sfschen/velocileptors}{\faGithub}.\footnote{\url{https://github.com/sfschen/velocileptors}} These codes use slightly different formalisms and are difficult to compare directly with each other. Complementing the recent works which performed detailed comparison of PyBird and {\sc CLASS-PT} codes when applied to data from BOSS survey \cite{Simon:2022lde,Simon:2022adh,Holm:2023laa}, having additional codes to compare with is still a useful feature. Additionally, to perform parameter estimation, some of these codes require python wrappers to Einstein-Boltzmann codes or external pipelines. In this paper, we use for the first time a new implementation of the redshift-space galaxy power spectrum calculation called {\sc CLASS-OneLoop}. This calculation is implemented in a native and flexible way within the {\sc CLASS} Boltzmann solver, and will be released publicly in the near future \cite{release}. Further technical details about the numerical implementation and optimisation strategies of the code will be described in a forthcoming release paper. The main goal of this work is to show that {\sc CLASS-OneLoop} compares well with other existing public codes, and that it provides consistent result when fitting data from N-body simulations or mock power spectra in sensitivity forecasts.

The rest of the paper is organized as follows; In section~\ref{sec:th}, we describe the theoretical model implemented in {\sc CLASS-OneLoop} for the calculation of the non-linear power spectrum of biased tracers in redshift space. We also briefly summarise the performance of {\sc CLASS-OneLoop} and we compare its predictions to those of CLASS-PT. In section~\ref{sec:sims}, as a first application of the code, we present a likelihood analysis of the {\sc Abacus} N-body simulations, and in section~\ref{sec:forecasts}, we present an \texttt{MCMC} forecast for a stage-IV (DESI-like) galaxy survey. Finally, in section~\ref{sec:conc}, we draw our conclusions. In appendix~\ref{app:SPM}, we describe a simpler phenomenological model (SPM) for the calculation of the galaxy power spectrum, which has been used frequently in the previous literature and that we employ as a comparison point. In appendix~\ref{app:FFTLog}, we provide technical details on the \texttt{FFTLog} method implemented in {\sc CLASS-OneLoop}. In appendix~\ref{app:loops}, we show our notations for the list of terms contributing to the one-loop galaxy power spectrum, and in appendix~\ref{app:nuisance} we show additional plots from our fits to {\sc Abacus} mocks and sensitivity forecast for a stage-IV survey to further illustrate parameter degeneracies.

\section{Redshift-space galaxy power spectrum: modeling non-linearities}\label{sec:th}
In this section, we describe the theoretical model of the one-loop halo or galaxy clustering power spectrum implemented in the {\sc CLASS-OneLoop} code. We don't aim at reviewing the large body of existing work on this topic, and focus instead on the main ingredients of our implementation of Eulerian perturbation theory in the {\sc CLASS-OneLoop} code, which follows closely ref.~\cite{Chen:2020fxs} by Chen, Vlah, and White. 

\subsection{Theoretical ingredients}\label{subsec:EFT}

We adopt the model of Chen et al. \cite{Chen:2020fxs} for the redshift-space one-loop halo power spectrum, based on an Eulerian moment expansion framework (also referred to as distribution function model \cite{Seljak:2011tx,Vlah:2012ni,Vlah:2013lia}) to describe the nonlinear redshift-space distortions, and using Eulerian perturbation theory to compute the loop contributions. We include halo biases up to the third order and account for the relevant EFT counter terms, stochastic contributions, infrared (IR) resummation, and Alcock-Paczynski (AP). Below we summarize the main ingredients for the model and refer the interested reader to \cite{Chen:2020fxs} for further details. 

\subsubsection{Perturbative modeling of nonlinearities}

The galaxy power spectrum expanded up to one-loop order in perturbation theory receives contributions from galaxy biases up to third-order in perturbation theory. At this order, assuming Gaussian initial conditions, the galaxy density field $\delta_g$ can be expanded in terms of the matter density field $\delta$, gravitational potential $\Phi$, velocity potential $\Phi_v$ and stochastic terms using the full renormalized bias expansion as \cite{Desjacques:2016bnm}
\begin{align}\label{eq:G_bias}
\delta_g(\bx)&=b_1\delta(\bx)+b_{\nabla^2\delta}\nabla^2\delta(\bx)+\epsilon(\bx) + \frac{b_2}{2} \delta^2(\bx)+ b_{\mathcal{G}_2}\mathcal{G}_2(\bx) + \epsilon_\delta(\bx) \delta(\bx)+\frac{b_3}{6} \delta^3(\bx) \nonumber \\
& +b_{\mathcal{G}_3}\mathcal{G}_3(\bx)+b_{(\mathcal{G}_2\delta)}\mathcal{G}_2(\bx)\delta(\bx) +b_{\Gamma_3}\Gamma_3(\bx)+ \epsilon_{\delta^2}(\bx) \delta^2(\bx)+ \epsilon_{\mathcal G_2}(\bx) {\mathcal G_2}(\bx), 
\end{align}
where we omit the explicit time dependence for convenience. The $\cG_2$ and $\cG_3$ are the second- and third-order Galileon operators
\begin{align}
\mathcal{G}_2(\Phi) & \equiv(\d_i\d_j\Phi)^2-(\d^2\Phi)^2, \\
\mathcal{G}_3(\Phi) & \equiv-\d_i\d_j\Phi\d_j\d_k\Phi\d_k\d_i\Phi-\frac{1}{2}(\d^2\Phi)^3 +\frac{3}{2}(\d_i\d_j\Phi)^2\d^2\Phi\,, 
\end{align}
while $\Gamma_3$ is the difference between density and velocity tidal tensors \cite{Chan:2012jj}, 
\begin{align}
\Gamma_3 &\equiv\mathcal{G}_2(\Phi)-\mathcal{G}_2(\Phi_v),
\end{align}
and the fields $\epsilon_{\mathcal O}$ refer to the stochastic contributions.

The galaxy positions measured in galaxy redshift surveys \cite{DESI:2016fyo, Amendola:2016saw, SPHEREx:2014bgr} are defined in redshift space. This implies that the line-of-sight velocity of galaxies contributes to their observed redshift, causing a Doppler shift in addition to the redshift caused by the cosmological expansion. Furthermore, the position of objects in redshift space, $\bs$, is shifted with respect to the one in real space, $\bx$, due to the peculiar velocity, $\bv$,  of halos/galaxies, such that $\bs = \bx + \hn \, (\bv.\hn)/\cH$, where $\hat n$ is the line of sight (LoS) direction, and $\cH = aH$ is the conformal Hubble parameter. This velocity-induced mapping from real- to redshift-space introduces an anisotropy in the clustering pattern, referred to as redshift-space distortions (RSD). On the one hand, RSD introduce an additional challenge in modeling the clustering statistics, on the other hand, they can be used to test the theory and probe the growth of large-scale structure \cite{Weinberg:2013agg}. 

The galaxy densities in real and redshift spaces, denoted as $\delta_g(\bx)$ and $\delta_g^s(\bs)$ respectively, are related to each other as
\begin{align}
    1 +\delta_g^s(\bs) &= \int d^3 x \left(1+\delta_g(\bx) \right) \delta_D(\bs-\bx-\bu), \notag \\
    (2\pi)^3 \delta_D(\bk) + \delta_g^s(\bk) &= \int d^3 x \left(1+\delta_g(\bx) \right) e^{i\bk.(\bx+\bu(\bx))},
\end{align}
where we dropped the explicit time dependence of the field and defined $\bu = \hn \, (\bv.\hn)/\cH$.  Defining the wavevector component along the LoS as $k_\parallel = k \mu$ 
and the coordinate shift as a field $\bu = u(\bx) \, \hat{n}$, we can write the Fourier transform of $u(\bx)$ in terms of the velocity divergence $\theta$ as  $u(\bk) = i k_\parallel/(\cH k^2)\theta(\bk)$. At linear order in PT, the velocity divergence is related to density as $\theta(\bk) = - f \cH \delta(\bk)$.  The power spectrum in redshift space can then be obtained from the velocity moment-generating function as
\be\label{eq:ps_master}
    P_g^s(\bk) = \int d^3 r \langle (1+\delta_g(\bx_1)(1+\delta_g(\bx_2))e^{i\bk.\Delta \bu}\rangle_{\bx_1 - \bx_2 = \br}\, ,
\ee
where $\Delta \bu = \bu(\bx_1) - \bu(\bx_2)$ is the pairwise velocity. To compute the redshift-space power spectrum at a given loop order, one needs to perform a series expansion of the exponent in eq. \eqref{eq:ps_master}. The moment expansion approach implies an expansion of the form
\begin{empheq}[box=\exactfbox]{align}\label{eq:rsd_ps}
    P_g^s(k,\mu) = \Sigma_{\hn}^{(0)} + i k \mu \Sigma_\hn^{(1)} + \frac{i^2}{2} k^2 \mu^2 \Sigma_\hn^{(2)} + \frac{i^3}{3!} k^3 \mu^3 \Sigma_\hn^{(3)} + \frac{i^4}{4!} k^4 \mu^4 \Sigma_\hn^{(4)}, 
\end{empheq}
where $\mu = \hat k.\hat n$ is the cosine of the angle with respect to the LoS, and $\Sigma_\hn^{(n)} = \langle(1+\delta_1)(1+\delta_2) \Delta \bu_1 ... \Delta \bu_n \rangle$ are density-weighted pairwise velocity moments, which include stochastic and EFT counter terms in addition to deterministic terms. We have only kept contributions of the moments containing up to 4 velocities. The first moments appearing in eq. \eqref{eq:ps_master} are given by 
\begin{align}\label{eq:Sigma} 
    \Sigma^{(0)}_\hn(k,\mu) &= P_{0 0}(k) + c_0^{(0)} k^2 P_0(k) + P_{\rm shot}^{(0)}, \\
    \Sigma^{(1)}_\hn(k,\mu)  &= 2 P_{01}(k,\mu) - i c_1^{(0)} f \mu k P_0(k) + ... \,,  \label{eq:mom1} \\
    \Sigma^{(2)}_\hn(k,\mu)  &= 2 P_{02}(k,\mu)- 2P_{11}(k,\mu)- 2 f^2 \left(c_2^{(0)} + c_2^{(2)}\mu^2\right) P_0(k) + P_{\rm shot}^{(2)} + ... \, , \label{eq:mom2} \\
    \Sigma^{(3)}_\hn(k,\mu)  &= 2 P_{03}(k,\mu) - 6P_{12}(k,\mu)+ 6i f^3 \left(c_3^{(0)} + c_3^{(2)}  \mu^2\right) \frac{\mu}{k}P_0(k) + ...\, , \\
    \Sigma^{(4)}_\hn(k,\mu)  &= - 8 P_{13}(k,\mu) + 6 P_{22}(k,\mu) + 24 f^4 c_4^{(2)}\frac{\mu^2}{k^2}P_0(k) + P_{\rm shot}^{(4)} + ... \, ,
\end{align}
where $P_{\rm shot}^{(n)}$ are the stochastic terms and $c_n^{(m)}$ the EFT counter terms for the $n$-th velocity moment $\Sigma_{\hat n}^{(n)}$. The superscript $(m)$ of the counterterms refer to the fact that terms with different dependence on the LoS direction can have different counterterm. The $P_{ij}$ stand for the cross correlations of different velocity moments,
\be
P_{ij}(k,\mu)\equiv \langle [(1+\delta_g) u^i](\bk) [(1+\delta_g)u^j](\bk') \rangle'. 
\ee
The prime in $\langle ... \rangle'$ denotes the expectation values without the factor of $(2\pi)^3\delta(\bk+\bk')$. The square bracket represents a convolution of two fields, and the indices $i,j$ refer to the number of velocity fields appearing in the cross correlation. More explicitly, we have
\begin{align}\label{eq:pnn_loops}
    P_{00}(k) &= \langle \delta_g(\bk')\delta_g(\bk)\rangle', \\
    P_{01}(k,\mu) &= \langle \delta_g(\bk') u(\bk)\rangle + \langle \delta_g(\bk') \left[\delta_g u \right](\bk)\rangle',   \\ 
    P_{02}(k,\mu) &= \langle \delta_g(\bk') \left[u^2 \right](\bk)\rangle + \langle \delta_g(\bk') \left[\delta_g u^2 \right](\bk)\rangle', \\
    P_{11}(k,\mu) &= \langle u(\bk') u(\bk)\rangle + 2\langle u(\bk') \left[\delta_g u \right](\bk)\rangle + \langle \left[\delta_g u\right](\bk') \left[\delta_g u\right](\bk)\rangle',   \\ 
    P_{03}(k,\mu)&= \langle \delta_g(\bk') \left[u^3 \right](\bk)\rangle', \\
    P_{12}(k,\mu)&=  \langle u(\bk') [u^2](\bk)\rangle + \langle u(\bk') \left[\delta_g u^2 \right](\bk)\rangle + \langle \left[\delta_g u\right](\bk') [u^2] (\bk)\rangle',\\
    P_{13}(k,\mu)&= \langle u(\bk') \left[u^3 \right](\bk)\rangle',\\
    P_{22}(k,\mu) &= \langle \left[u^2 \right](\bk') \left[u^2 \right](\bk)\rangle'. 
\end{align}
The explicit expressions of the $P_{ij}$ are given in appendix \ref{app:loops} and Ref. \cite{Chen:2020fxs}. The key point to note is that the form of the loop integrals is such that we can separate the dependence of the loops on the LoS direction in terms of a finite series in $\mu$, and evaluate the loops using FFTLog independent of the LoS direction.\footnote{As we will briefly discuss in section~\ref{sec:IR}, the implementation of the IR resummation breaks the above assumption, and thus we make the approximation in eq. \eqref{eq:IR_redshift_approx} to achieve the separation of LoS dependence and write the loop integrals in terms of integrals over the isotropic power spectrum and pull the anisotropic suppression outside of the loop integrals.} Collecting the stochastic and counter term contributions in eq. \eqref{eq:ps_master}, we have
\begin{align}
P_{\rm shot}(k,\mu) &= \frac{1}{\bar n} \left[1 + s_0 + s_1 k^2 + s_2 f^2 \mu^2 k^2 + s_3 f^4 \mu^4 k^4 \right], 
\label{eq:expansion_shot}\\
P_{\rm ct}(k,\mu) &= \left\{c_0^{(0)}+ \left[c_1^{(0)} + f c_2^{(0)}\right] f\mu^2 + \left[c_2^{(2)} + f c_3^{(0)}\right] f^2\mu^4 + \left[c_3^{(2)} + f c_4^{(2)}\right] f^3 \mu^6 \right\} k^2 P_0(k). 
\label{eq:expansion_ct_full}
\end{align}
In writing the expression for the shot noise, we have Taylor expanded the stochastic contribution to the real-space density field to include the leading scale-dependence, i.e. $P_\epsilon(k) = s_0 + s_1 k^2$. For stochastic contributions to higher-order velocity moments, we have only kept the constant piece in the Taylor expansion. Regarding the counter-terms, the impact of different terms multiplying the same power of $\mu$ are fully degenerate when fitting data. Thus, in both our fits to {\sc Abacus} simulations and forecasts, we vary a single parameter for each power of $\mu$. Therefore, we consider
\be
P_{\rm ct}(k,\mu) = \left[c_0 + c_1 f\mu^2 + c_2 f^2\mu^4 + c_3 f^3 \mu^6 \right] k^2 P_0(k). 
\label{eq:expansion_ct}
\ee
The counter terms are in principle redshift dependent. When fitting survey data, one should vary a set of counter terms for each redshift bin. However, in our forecasts, we neglect this redshift dependence and vary a single parameter for each of the counter-terms across all redshift bins, as described in section~\ref{sec:forecasts}. We will see that even with redshift-independent counter terms our recipe has enough freedom to extract unbiased information from simulated data.

\subsubsection{IR resummation}\label{sec:IR}

Finally, one needs to take into account the fact that large-scale bulk flows induce large displacements in the matter density field on comoving scales of order $\sim 10 \ {\rm Mpc}$. While these bulk flows are unobservable locally due to equivalence principle \cite{Peloso:2013zw,Kehagias:2013yd,Creminelli:2013mca}, they smooth features (such as BAO wiggles) in the power spectrum. In standard perturbation theory, the effect of bulk flows is treated perturbatively, which is not optimal to describe large displacements, and thus the predicted shape of BAOs has a limited accuracy \cite{Eisenstein:2006nj,Crocce:2007dt,Sugiyama:2013gza}. Instead, in Lagrangian Perturbation Theory (LPT) \cite{Matsubara:2007wj,Carlson:2012bu,Porto:2013qua,Vlah:2015sea}, the treatment of bulk flows is non-perturbative since the contribution arising from the (linear) displacement field can be resummed. A similar resummation can be performed in SPT \cite{Baldauf:2015xfa,Vlah:2015zda}, in a hybrid LPT-SPT approach \cite{Senatore:2014via,Senatore:2017pbn,Lewandowski:2018ywf}, or within Time-Sliced Perturbation Theory \cite{Blas:2016sfa,Ivanov:2018gjr}. 

Given that displacements only affect the BAO wiggles, a practical approach to implement the resummation in real space consists in splitting the linear matter power spectrum into a wiggle and a no-wiggle (smooth) component, 
\be
P_0(k)= P_{\text{nw}}(k)+P_{\text{w}}(k)\,,
\ee
and to apply a damping factor to the wiggle part to obtain the leading-order IR-resummed matter power spectrum,
\be\label{eq:LO_real}
P_{\text{LO}}(k) \equiv P_{\text{nw}}(k)+e^{-k^2\Sigma^2}P_{\text{w}}(k),
\ee
with LO standing for leading order.
The damping exponent is given by
\be
\Sigma^2 \equiv\frac{1}{6\pi^2}\int_0^{k_s}dq \ P_{\text{nw}}(q) \left[1-j_0\left(\frac{q}{k_{\rm osc}}\right)+2j_2\left(\frac{q}{k_{\rm osc}}\right)\right]\,,
\label{eq:Sigma_integral}
\ee
where $k_{\rm osc}$ is the inverse of the BAO scale, $k_{\rm osc} \sim 110\ {\rm Mpc}/h$, $k_s$ is the separation scale controlling the modes to be resummed, and $j_n$ are the spherical Bessel function of order $n$.  At next-to-leading order (NLO), the matter power spectrum can be obtained by using the expression in eq.~\eqref{eq:LO_real} as an input in the expression of the one-loop matter power spectrum,
\be \label{eq:IR_real}
P_{\text{NLO}}(k) \equiv P_{\text{nw}}(k) +e^{-k^2\Sigma^2}P_{\text{w}}(k)(1+k^2\Sigma^2) + P_{\text{loop}}[P_0\rightarrow P_{\text{LO}}](k)  \,,
\ee
where $P_{\text{loop}}$ is considered a function of the linear power spectrum. For a biased tracer, the computation is equivalent, with only the linear bias parameter multiplying the first two terms, and the loop contributions due to non-linear biases computed with the LO power spectrum as input instead of the linear power spectrum. 

In principle, $k_s$ is arbitrary, and any dependence on it should be treated as a theoretical error. 
However, it has been shown that the exact value of $k_s$ can affect the leading-order IR-resummed power spectrum, while  values of $k_s$ in the range of $(0.05-0.2)$ are indistinguishable at the level of the next-to-leading order spectrum~\cite{MoradinezhadDizgah:2020whw}. Thus, in our approach, we keep this parameter fixed to $k_s = 0.2\ {\rm Mpc}^{-1}h$.

Several algorithms have been used in the literature to compute the no-wiggle spectrum $P_\mathrm{nw}(k)$ for a given linear spectrum $P_0(k)$ -- a step called broadband extraction. This includes the semi-analytic formula of ref. \cite{Eisenstein:1997ik,Eisenstein:1997jh,Kiakotou:2007pz}, a Bspline-based approach with fixed location of the nodes of the splines \cite{Reid:2009xm}, a fourth-order polynomial fit to the linear spectrum \cite{Hamann:2010pw}, the Savitzky-Golay filtering \cite{boyle2018deconstructing, savitzky1964smoothing}, the discrete spectral method (DST) \cite{Hamann:2010pw, Baumann:2017gkg}, smoothing with a Gaussian filter (Gfilter), and Bspline-basis regression \cite{Vlah:2015zda}. A summary of the last three methods and a detailed comparison of the corresponding IR-resumed power spectra can be found in Refs. \cite{Vlah:2015zda,MoradinezhadDizgah:2020whw}. Here, we use the Gfilter method as a baseline. Clearly, the splitting of the power spectrum into wiggle and no-wiggle contributions is not unique. Nevertheless, one expects that different splitting methods should result in consistent predictions of the IR-resummed matter and galaxy power spectra. In section~\ref{sec:code_validation}, we discuss the impact of the splitting method on the IR-resummed one-loop galaxy power spectrum in real and redshift spaces by comparing the predictions using DST and Gfilter algorithms.\footnote{See also ref.~\cite{MoradinezhadDizgah:2020whw} for a comparison of the DST, Gfilter, and Bspline methods at the level of the galaxy power spectrum in real space.}

In redshift space, in addition to resumming the long-wavelength displacements, one can also resum the long-wavelength velocity modes, both of which only affect the BAO wiggles. At leading-order, the IR-resummed matter power spectrum is given by 
\be
P^s_{\rm LO}(k,\mu) = (1+f\mu^2)^2 \left[P_{\text{nw}}(k)+e^{-k^2\Sigma_{\rm s}^2(\mu)}P_{\text{w}}(k)\right].
\ee
The damping exponent now depends on the angle w.r.t. the line of sight direction,
\be\label{eq:rsd_damp}
\Sigma_{\rm s}^2(\mu) = \left[1+f(f+2)\mu^2 \right] \Sigma^2 + f^2 \mu^2(\mu^2 - 1)\delta \Sigma^2, 
\ee
where the new redshift-space contribution $\delta \Sigma$ is given by
\be
\delta \Sigma \equiv \frac{1}{2\pi^2} \int_0^{k_s} dq \ P_{\rm nw}(q) \ j_2\left(\frac{q}{k_{\rm osc}}\right). 
\ee
Computing the IR-resummed loops in redshift space is more intricate than in real space because the anisotropic damping of BAOs renders the loop integrals 2-dimensional. In analogy to real-space computation, the IR-resummed power spectrum in redshift-space is given by \cite{Ivanov:2018gjr}
\be\label{eq:IR_redshift}
    P^{s, {\rm IR}}_g(k,\mu) = (b_1+f\mu^2)^2 \left[ P_{\text{nw}}(k) +e^{-k^2\Sigma_{\rm s}^2(\mu)}P_{\text{w}}(k)\left(1+k^2\Sigma_{\rm s}^2(\mu)\right) \right] + P^s_{g,\text{loop}}[P_0\rightarrow P_{\rm LO}](k,\mu),
\ee
where in the last term, the linear matter power spectra in the loop integrals, $P_0$, are replaced by the leading-order IR-resummed matter power spectra, $P_{\rm LO}$. The fact that the exponential suppression ${\rm exp}[-k^2\Sigma^2(\mu)]$ applied to the wiggle contribution in $P_{\rm LO}$ is anisotropic complicates the computation of the loops since we can not anymore separate the dependence of the loops on the LoS as finite series of $\mu$. Neglecting the loop contributions involving two insertions of $P_{\rm w}$, a sufficiently accurate approximation to overcome this challenge is to split the loop contributions into smooth and wiggly parts and apply the exponential suppression on the wiggly part. Therefore, one needs to compute the loops twice, once with two insertions of the no-wiggle component, and the second time with one insertion of the wiggle component and one of the no-wiggle component. Then we apply the direction-dependent damping on the latter one \cite{Ivanov:2018gjr, Chudaykin:2020aoj}.

In practice, to ensure the accuracy of the FFTLog computation of the loops, we perform the IR resummation by applying the suppression factor on the difference between loops computed with the linear matter power spectrum and the no-wiggle spectrum and add it to the loops computed with the no-wiggle spectrum. Therefore, the IR-resummed one-loop galaxy power spectrum in redshift space is given approximately by
\begin{align}\label{eq:IR_redshift_approx}
    P^{s, {\rm IR}}_g(k,\mu) &\simeq (b_1+f\mu^2)^2 \left[ P_{\text{nw}}(k) +e^{-k^2\Sigma_{\rm s}^2(\mu)}P_{\text{w}}(k)\left(1+k^2\Sigma_{\rm s}^2(\mu)\right) \right]\notag \\
    &+ P^s_{g,\text{loop}}[P_{\rm nw}](k,\mu) + e^{-k^2\Sigma_{\rm s}^2(\mu)} \left\{P^s_{g,\text{loop}}[P_0](k,\mu)-P^s_{g,\text{loop}}[P_\mathrm{nw}](k,\mu)\right\}\,,
\end{align}
We note that this expression includes contributions with two insertions of $P_{\rm w}$ in the first term in the curly bracket that should be suppressed by two powers of the exponential suppression, instead of one. This approximation, however, should have negligible impact on the accuracy of the model (see section \ref{sec:code_validation} for further discussion about the consequence of this approximation).   

\subsubsection{Alcock-Paczynski effect}\label{subsub:AP}

When analyzing observational data, one assumes a fiducial cosmology to convert the observed angular position and redshift of each galaxy to comoving distances in order to compute the 3D power spectrum. If this fiducial cosmology differs from the truth, the inferred radial and transverse distance (defined with respect to the line-of-sight) are distorted differently. These distortions render the observed statistics anisotropic, an imprint referred to as the Alcock-Paczynski (AP) effect \cite{Alcock:1979mp}. The AP effect on the halo or galaxy power spectrum can be modeled by applying an overall volume re-scaling factor to the galaxy power spectrum computed in a given redshift bin, while remapping the Fourier wave vectors of the fiducial cosmology, $\bk_{\rm fid}$, into wave vectors in the true cosmology, $\bk_{\rm true}$. 

Therefore, including the AP effect, the theoretical predictions for the observed power spectrum in terms of wave-vectors in the fiducial and true cosmologies are related as
\be\label{eq:Pkmu_AP}
P_{g,\rm AP}^{s,\rm IR}(\bk_{\rm fid}) = \alpha_\parallel^{-1} \alpha_\perp^{-2} P^{s,{\rm IR}}_g(\bk_{\rm true}(\bk_{\rm fid})),
\ee
where the the two wave vectors have their radial and transverse parts related through $k_{\parallel,\perp}^{\rm fid} = \alpha_{\parallel, \perp}k^{\rm true}_{\parallel, \perp}$, and the AP parameters, $\alpha_\parallel$ and $\alpha_\perp$, are given by \cite{Alcock:1979mp,Padmanabhan:2008ag}
\be
\alpha_\parallel \equiv  \frac{H^{\rm fid}(z)}{H(z)}, \qquad \qquad \alpha_\perp \equiv \frac{D_A(z)}{D_A^{\rm fid}(z)}.
\ee

\subsubsection{Power spectrum wedges and multipoles}

The commonly applied method to extract the cosmological information from the measured anisotropic galaxy power spectrum is to project the dependence on the LoS direction into Legendre multipole moments \cite{Padmanabhan:2008ag}. Including the AP effect, the theoretical prediction of power spectrum multipoles is given by 
\be
P_\ell(k_{\rm fid}) = \frac{2\ell+1}{2}\int_{-1}^1 d\mu_{\rm fid} \ P_{g,\rm AP}^{s,\rm IR}(k_{\rm fid},\mu_{\rm fid}) \ {\mathcal L}_\ell(\mu_{\rm fid}),
\ee
where $P_{g,\rm AP}^{s,\rm IR}(k_{\rm fid},\mu_{\rm fid})$ is the 2D power spectrum in terms of the wavevectors in the fiducial cosmology and is given in eq. \eqref{eq:Pkmu_AP}, and the ${\cal L}_\ell(x)$'s are  Legendre multipoles.

Alternatively, one can define power spectrum wedges, which correspond to the averaged power spectrum over wide angular bins of $\mu$ \cite{Kazin:2011xt,Grieb:2015bia}and is given by
\be\label{eq:wedges}
P(k_{\rm fid},\mu_i) = \frac{1}{\Delta \mu}\int_{\mu_i-\Delta \mu/2}^{\mu_i+\Delta \mu/2} d\mu_{\rm fid} \  P_{g,\rm AP}^{s,\rm IR}(k_{\rm fid},\mu_{\rm fid}),
\ee
where $\mu_i$ is the center of an angular bin of width $\Delta \mu$, and the angular bins have lower and higher edges of $[-1,1]$. Since the 2D power spectrum is symmetric under the exchange of $\mu\rightarrow-\mu$, the wedges are commonly defined in the range of $\mu \in [0,1]$. 

The 2D power spectrum $P(k,\mu)$ is a relatively smooth function of $\mu$ and its amplitude is dominated by the monopole and quadrupole terms. Therefore, one expects that the choice of the wedges or multipoles as observables are roughly equivalent and provide comparable cosmological constraints. In practice, however, this choice gives rise to subtle differences \cite{Chen:2020fxs}.

The power spectrum wedges offer two advantages over multipoles; first, one can effectively isolate the dominant contribution of the FoG effect \cite{Chen:2020fxs,Chudaykin:2019ock}, and second, the contamination of the measurements due to a number of observational systematics \cite{Hand:2017pqn} is localized in a subset of the wedges. The former is possible since the effect of strong FoG on wedges are primarily limited to angular bins with $|\mu|\simeq1$. Thus, discarding these bins effectively removes the dominant FoG effect while retaining most of the information of the anisotropic power spectrum. The residual theoretical error of the wedges is dominated by the two-loop corrections to the matter power spectrum \cite{Chudaykin:2019ock}. In contrast, strong FoG impact all multipoles beyond the monopole and limit the validity regime of the perturbative description of the multipoles. Commonly, in the analysis of the multipoles, a considerably lower scale cut, $k_{\rm max}$, is imposed for the quadrupole, which results in an information loss for the contribution of the lower $\mu$ bins to the quadrupole.\footnote{Potentially, one can alleviate this loss by down-weighting the contribution of high $\mu$ modes by constructing appropriate scale-dependent weights for the multipoles which reduce the contamination by FoG effect \cite{Chen:2020fxs}.} 

Regarding observational systematics, redshift errors affect the $\mu\simeq 1$ bins, while systematics such as fiber collisions predominantly affect transverse Fourier modes. Thus, their effect is localized in $\mu\simeq0$ wedges. Therefore, removing the purely transverse and radial modes can considerably reduce the contamination of the measured wedges by observational systematics \cite{Hand:2017pqn}. Like FoG effects, observational systematics affect all multipoles. However, in this case, a possible strategy can be devised to isolate the most contaminated modes by down-weighting their contributions to the multipoles \cite{Hand:2017pqn}. 

In this work, we adopt a theoretical model of the redshift-space galaxy power spectrum based on velocity moment expansion \cite{Chen:2020fxs}, which is an expansion in $k$ and $\mu$.
As such, it naturally describes the power spectrum wedges with well-defined convergence properties. A detailed comparison of the power spectrum wedges and multipoles predicted by this model against simulated halo catalogs can be found in ref. \cite{Chen:2020fxs}. In all the analyses presented in this paper, we adopt the power spectrum wedges as our observables, even though {\sc CLASS-OneLoop} can also compute power spectrum multipoles.

\subsection{Numerical evaluation with CLASS}

In this section, we provide an overview of our implementation of the redshift-space galaxy power spectrum calculation within the {\sc CLASS} Boltzmann code. We anticipate making the code publicly available in an upcoming release of the master {\sc CLASS} branch once we have completed final testing and additional optimization. The presented code is built upon earlier implementations of the one-loop real-space EFT power spectrum in two codes: {\sc CosmoSIS-gClust}\footnote{The new version of this code, including the redshift-space distortions, will also become publicly available soon.}, used in refs. \cite{MoradinezhadDizgah:2020whw,EuclidPTchallenge}, and {\sc LimHaloPT} \cite{limHaloPT} \href{https://github.com/amoradinejad/limHaloPT}{\faGithub}\footnote{\url{https://github.com/amoradinejad/limHaloPT}}, used in ref. \cite{MoradinezhadDizgah:2021dei}. To comprehensively validate the code, we conducted a parallel implementation in the {\sc CosmoSIS-gClust} code. Despite several differences at the level of the numerical implementation, particularly in the utilization of external libraries, the two codes yield excellent agreement. The new implementation of the galaxy power spectrum within {\sc CLASS} offers several noteworthy advantages. It is seamlessly integrated into the master branch of {\sc CLASS}, which ensures continuous maintenance and ease of accessibility within various likelihood analysis tools, including 
{\sc MontePython} \cite{Audren:2012wb, Brinckmann:2018cvx}
\href{https://github.com/brinckmann/montepython_public}{\faGithub}\footnote{\url{https://github.com/brinckmann/montepython_public}}, 
{\sc Cobaya} \cite{Torrado:2020dgo}
\href{https://github.com/CobayaSampler/cobaya}{\faGithub}\footnote{\url{https://github.com/CobayaSampler/cobaya}},
and {\sc CosmoSIS} 
\cite{Zuntz:2014csq} 
\href{https://github.com/joezuntz/cosmosis}{\faGithub}\footnote{\url{https://cosmosis.readthedocs.io/en/latest/}}. This full integration allows to avoid redundancies, duplicate tasks and parameter mappings across codes, interfaces and wrappers, which are a frequent source of error. The style of the {\sc CLASS-OneLoop} module is homogeneous with the rest of the {\sc CLASS} code and equally documented. The new module can be selected with a switch that also offers alternative methods for the calculation of the non-linear galaxy power spectrum in real space such as {\sc HALOFIT} \cite{Takahashi:2019hth} and {\sc HMCode} \cite{Mead:2020vgs}, which can be readily compared to one-loop EFT calculation. Finally, the new code is as easy to compile as previous versions of {\sc CLASS} since the C code does not rely on any external C libraries\footnote{This is true at least when the code is used in its default ``{\tt FFTLog}'' mode. At the time of writing, when the code is used in its (slow) ``direct integration'' mode for the purpose of comparison and validation, it still uses one external C library, as explained in \S~\ref{sec:imp_perf}.} and the {\tt classy} python wrapper does not require additional python modules compared to previous versions.

As a second reference for the cross-validation of {\sc CLASS-OneLoop}, we carried out detailed comparisons with the predictions from the {\sc CLASS-PT} code, the results of which we present in this section. The theoretical models implemented in the two codes are nearly identical, apart from the choice of some of the EFT counter terms and stochastic contributions. Both codes use the same Eulerian bias expansion basis, which makes the comparison more straightforward. We have also compared the {\sc CLASS-OneLoop} output with Eulerian-EFT calculations from {\sc velocileptor} \cite{Chen:2020fxs,Chen:2020zjt} \href{https://github.com/sfschen/velocileptors}{\faGithub}\footnote{\url{https://github.com/sfschen/velocileptors}}. We found a good agreement between the two codes. However, the difference in the bias expansion assumed in the two implementations makes the direct comparison less straightforward. Therefore, we do not show the detailed comparison here.

\subsubsection{Current implementation}
\label{sec:imp_perf}

The code can be run in two modes that define the way in which the loop integrals are computed: direct integration (DI) or evaluation using an \texttt{FFTLog} approach. To check that the \texttt{FFTLog} method is implemented in a robust and accurate way, despite of several choices and subtleties inherent to this approach~\cite{Hamilton:1999uv}, we use the direct integration method as a robust reference for a term-by-term comparison of the loop integrals.

For direct integration, we employ the \textsc{CUBA} library \cite{Hahn:2004fe,Hahn:2014fua} \href{https://feynarts.de/cuba/}{\faGlobe} \footnote{\url{https://feynarts.de/cuba/}}, primarily using the \texttt{Cuhre} routine, which utilizes the cubature rule to perform deterministic multi-dimensional integration. Other \textsc{CUBA} routines could easily be used instead of \texttt{Cuhre}, but we have observed the fastest convergence with \texttt{Cuhre}. Our \texttt{FFTLog} implementation follows the method outlined in ref. \cite{Simonovic:2017mhp} (and extended to redshift space in \cite{Chudaykin:2020aoj}). We summarize the basics of the latter method below and then describe its practical implementation within the code. 

The loop integrals summarised in section
\ref{subsec:EFT} are formally identical to those of a massless quantum field theory. This observation inspired the approach outlined in ref. \cite{Simonovic:2017mhp}, which relies on expanding each integrand in terms of powers of $q$ and $|\bk-\bq|$ such that they reduce to analytic expressions of the form
\begin{equation}
\label{eq:loops_fft}
\int_\bq \frac{1}{q^{2\nu_1} \kmq^{2\nu_2}} \equiv k^{3-2\nu_{12}} I(\nu_1,\nu_2),
\end{equation}
where $\int_\bq \equiv \int \frac{d^3 q}{(2\pi)^3}$, $\nu_{12} = \nu_1 + \nu_2$ and the function $I(\nu_1,\nu_2)$ is given in terms of Euler gamma functions,
\begin{equation}\label{eq:I-integral}
I(\nu_1,\nu_2) = \frac{1}{8 \pi^{3/2}} \frac{\Gamma(\frac{3}{2}-\nu_1)\Gamma(\frac{3}{2}-\nu_2)\Gamma(\nu_{12}-\frac{3}{2})}{\Gamma(\nu_1)\Gamma(\nu_2)\Gamma(3-\nu_{12})}.
\end{equation}
To cast the calculation in this form, one needs to expand each SPT kernel (e.g., $F_2^2(\bq,\kmq)$) and each nonlinear bias operator in powers of wavenumbers. Furthermore, one expands the matter power spectrum as a Fourier series in log-space. We first discretize the wavenumber space with
$(N_\mathrm{FFT}+1)$ values
$k_n=k_\mathrm{min}\,e^{n\Delta}$
ranging from 
$k_0=k_\mathrm{min}$
to
$k_{N_\mathrm{FFT}} = k_\mathrm{max}=k_\mathrm{min}\,e^{N_\mathrm{FFT}\Delta}$. 
The \texttt{FFTLog} expansion of the array of $(N_\mathrm{FFT}+1)$ values $P_0(k_n)$ reads
\begin{equation}
P_0(k_n) = \sum_{m=-N_\mathrm{FFT}/2}^{N_\mathrm{FFT}/2} c_m k_n^{\nu+i\eta_m},
\end{equation}
where the parameter $\nu$ is called the FFT bias, the index $\eta_m$ is defined as $\eta_m = m\times 2\pi/(N_\mathrm{FFT}\Delta)$, and the coefficients $c_m$ are given by the inverse transformation
\begin{equation}
c_m = \frac{k_0^{-i \eta_m}}{N_{\mathrm{FFT}}} \sum_{n=0}^{N_\mathrm{FFT}-1} \frac{P_0(k_n)}{k_n^\nu} e^{-2 \pi i m n / N_\mathrm{FFT}}~.
\end{equation}
Therefore, the linear matter power spectrum is biased by a power-law $k^\nu$.  This biased power spectrum may have divergent powers in the UV and/or IR limit, depending on the choice of $\nu$. If there is no value of the FFT bias such that the integrals converge in both limits, divergences have to subtracted by hand from the \texttt{FFTLog} calculated loops. 

Once the loop integrals have been recasted in the form of eq.~\eqref{eq:loops_fft}, they reduce to (numerically cheap) matrix or vector multiplications, where the matrices need to be computed once and for all for a given value of the FFT parameters  ($N_\mathrm{FFT}$, $\nu$, $k_{\rm min}$, $k_{\rm max}$). The cosmology dependence of the loops is captured by the coefficients $c_m$, which account for the input linear matter power spectrum describing a given cosmological model.

In redshift space, in addition to eq.~\eqref{eq:loops_fft}, there are four general forms for the loop integrals that have an explicit dependencies on the line-of-sight direction $\hat{z}$
\cite{Chudaykin:2020aoj},
\begin{align}\label{eq:mudep}
     \int_\bq \quad\frac{\hat{z} \cdot \bq}{q^{2\nu_1} \kmq^{2\nu_2}} &= k^{3-2(\nu_1+\nu_2)}k\mu A_1(\nu_1,\nu_2), \\
    \int_\bq \quad\frac{(\hat{z} \cdot \bq)^2}{q^{2\nu_1}\kmq^{2\nu_2}} &= k^{3-2(\nu_1+\nu_2)}k^2 \left[A_2(\nu_1,\nu_2)+\mu^2B_2(\nu_1,\nu_2)\right],\\
    \int_\bq \quad\frac{(\hat{z} \cdot \bq)^3}{q^{2\nu_1}\kmq^{2\nu_2}} &= k^{3-2(\nu_1+\nu_2)}k^3\mu \left[A_3(\nu_1,\nu_2)+\mu^2B_3(\nu_1,\nu_2)\right],\\
    \int_\bq \quad\frac{(\hat{z} \cdot \bq)^4}{q^{2\nu_1}\kmq^{2\nu_2}} &= k^{3-2(\nu_1+\nu_2)}k^4 \left[A_4(\nu_1,\nu_2)+\mu^2B_4(\nu_1,\nu_2)+\mu^4C_4(\nu_1,\nu_2)\right].\label{eq: LoS4}
\end{align}
The full analytical expression of these integrals can be found in Appendix~\ref{app:FFTLog}. 

The FFT parameters ($N_\mathrm{FFT}$, $\nu$, $k_{\rm min}$, $k_{\rm max}$) must be chosen with particular care. The \texttt{FFTLog} transformation involves a truncation of the matter power spectrum within a finite range $[k_\mathrm{min}, k_\mathrm{max}]$ and treats the truncated function as periodic in logarithmic space -- assuming periodic replications of it below $k_\mathrm{min}$ and above $k_\mathrm{max}$. These assumptions may cause ringing and aliasing effects \cite{Hamilton:1999uv}. To limit such artefacts, one should choose values of the bias and of the truncation interval such that the assumed periodic function does not feature a strong discontinuity at the edges, that is, such that $[P_0(k_\mathrm{min})/k_\mathrm{min}^\nu] \simeq 
[P_0(k_\mathrm{max})/k_\mathrm{max}^\nu]$. Ideally, for each cosmology, the algorithm could search for optimal value of the numbers ($\nu$, $k_\mathrm{min}$, $k_\mathrm{max}$) ensuring the continuity condition. However, this would be sub-optimal in the context of a parameter inference run, since the FFT matrices (that correspond to the FFT expansion of the  loop integration kernels) would need to be re-computed for each cosmological model. It is thus more efficient to fix the FFT parameters to values such that any cosmology with a power spectrum not too far from that of the $\Lambda$CDM {\texttt Planck2018} best-fit model gets only marginally affected by ringing and aliasing in the range relevant for fitting observations. As demonstrated in the following section, we achieve this target by fixing $\nu_m = -0.3$ for the calculation of
the total matter power spectrum and $\nu_g = -1.55$ for that of the galaxy (or more generally biased tracer) power spectrum. Then, in each of these cases, we set the expansion domain boundaries to $k_m\in [ 10^{-6}, 10^3] \ h/$Mpc for total matter and $k_g\in [ 10^{-6}, 4\times 10^3] \ h/$Mpc for galaxies. 

Note that using such a high $k_\mathrm{max}$ could increase the time needed by {\sc CLASS} to compute the linear power spectrum. This increase in computation time is not really worthwhile because, at the end, we are only interested in predictions for the matter or galaxy power spectrum on mildly non-linear scales, that is, up to $k\leq {\cal O}(1) \,h/\mathrm{Mpc}$. The calculation of the non-linear spectrum on such scales is very weakly affected by the precise value of the input linear spectrum at $k \geq {\cal O}(10^2)\, h/\mathrm{Mpc}$. This means that, on such large wavenumbers, an approximate input for the linear power spectrum is sufficient. Thus, to speed up the computation, we use an extrapolator of the linear matter power spectrum built within {\sc CLASS}. The extrapolation is of the form $P_0(k) = (a+b\,\ln k)^2k^{n_{\rm s}-4}$ and is also used by the {\sc HMCode} algorithm. We checked that, for a wide range of cosmological models, the non-linear power spectrum at $k\leq 1 h/\mathrm{Mpc}$ is robust against switching from an exact calculation of the linear power spectrum up to $k \sim 10^3 h/\mathrm{Mpc}$ to using the extrapolator at $k > 50\,$Mpc$^{-1}$. 

The choice of the size $N_\mathrm{FFT}$ of the FFT grid used for the expansion of the linear matter power spectrum affects the precision and the performance of the computation. A too small value of $N_\mathrm{FFT}$ would lead to spurious oscillations in the modelled power spectrum and compromise the precision of the method. In {\sc CLASS-OneLoop}, the default value is set to $N_{\rm FFT} = 256$, but we will also present in the next section some accuracy and timing estimates for higher values of $N_{\rm FFT}$.

Ultimately, the total matter or galaxy power spectra in real or redshift space are linear combinations of loop integrals weighted by input parameters such as biases, shot noise and counter-term coefficients, as well as powers of the angular variable $\mu$ in the case of the redshift-space power spectrum. Appendix \ref{app:loops} shows how the bias parameters $b_\alpha$ can be factorised; eq.~\eqref{eq:expansion_shot} shows the same for shot noise parameters $s_i$, \eqref{eq:expansion_ct} for counter-term parameters $c_j$, and \eqref{eq:mudep}-\eqref{eq: LoS4} for powers of $\mu$. Thus, it is efficient to compute and store each individual loop (independently of these parameters), and to combine them at the very end of the calculation for each given set of ($b_\alpha$, $s_i$, $c_j$, $\mu$). The user can decide if the final output of the {\sc CLASS-OneLoop} C code is a set of individual loop calculations, valid for any parameters ($b_\alpha$, $s_i$, $c_j$, $\mu$), or a final power spectrum for given values of these parameters. The individual loops are stored in a dedicated structure accessible to the python wrapper {\tt classy} and can also be assembled directly within the wrapper to form the power spectrum for multiple sets of parameters ($b_\alpha$, $s_i$, $c_j$, $\mu$), without re-running the C code.

This strategy has a decisive impact when using {\sc CLASS-OneLoop} within a parameter extraction pipeline like e.g., {\sc MontePython} or {\sc Cobaya}. In this case, the \texttt{FFTLog} matrices, which are independent of cosmology, can be computed once and for all during the the initialization phase of the MCMC run and remain saved in memory. For each new cosmological model, {\sc CLASS-OneLoop} computes the linear power spectrum and its FFT transform, does several matrix multiplications to evaluate each loop integral, and stores them in a dedicated structure. These few steps will usually dominate the total execution time of the pipeline, although these calculations remain very fast as we shall see in the next section. Then, at the level of the parameter extraction pipeline, a likelihood function accounting for a current or future experiment (e.g. a spectroscopic galaxy survey like eBOSS, DESI or {\it Euclid}) needs to evaluate the galaxy power spectrum for a given set of bias, shot noise and counter terms ($b_\alpha$, $s_i$, $c_j$) and for several angles (if the required observable is the redshift-space power spectrum $P(k,\mu,z)$) or several multipoles (if the observable is the Legendre-expanded spectrum $P_\ell(k,z)$). There is no need to re-run {\sc CLASS-OneLoop} for this. Dedicated functions in the {\tt classy} wrapper are designed to assemble the correct power spectrum for the required parameters ($b_\alpha$, $s_i$, $c_j$) and the required list of angles $\mu$ or multipoles $\ell$, with a negligible evaluation time.

Additionally, parameter extraction pipelines like e.g., {\sc MontePython} or {\sc Cobaya} make a distinction between fast (nuisance) parameters and slow (cosmological) parameters. The fast parameters are sampled more frequently for fixed sets of slow parameters, in order to save calls to the Boltzmann solver. This functionality is preserved by our implementation, with bias, shot noise and counter terms playing the role of fast (nuisance) parameters. Indeed, the spectrum can be obtained for different sets ($b_\alpha$, $s_i$, $c_j$) through repeated call to the the {\tt classy} wrapper, without any need to execute {\sc CLASS-OneLoop} again. This feature is essential to perform efficient MCMC runs when there is a plethora of bias, shot noise and counter term parameters to be marginalised over.

\subsubsection{Validation of FFTLog method versus direct integration}
\label{sec:code_validation_FFT_DI}

For our default value of $N_{\rm FFT} = 256$ and the above mentioned FFT domain and biases, we find an excellent agreement between the computations in DI and \texttt{FFTLog} mode. The individual contributions to the redshift-space galaxy spectra (given in appendix \ref{app:loops}) agree up to better than 0.75\% on large scales ($k<10^{-2}h/$Mpc), 0.1\% on intermediate scales, and 0.50\% on small scales ($k<0.3h/$Mpc). The total power spectrum obtained by combining all the loop contributions, each weighted by the corresponding biases, shows an even better agreement between the two methods.

\begin{figure}[htbp!]
    \centering
    \includegraphics[width=0.496\textwidth]{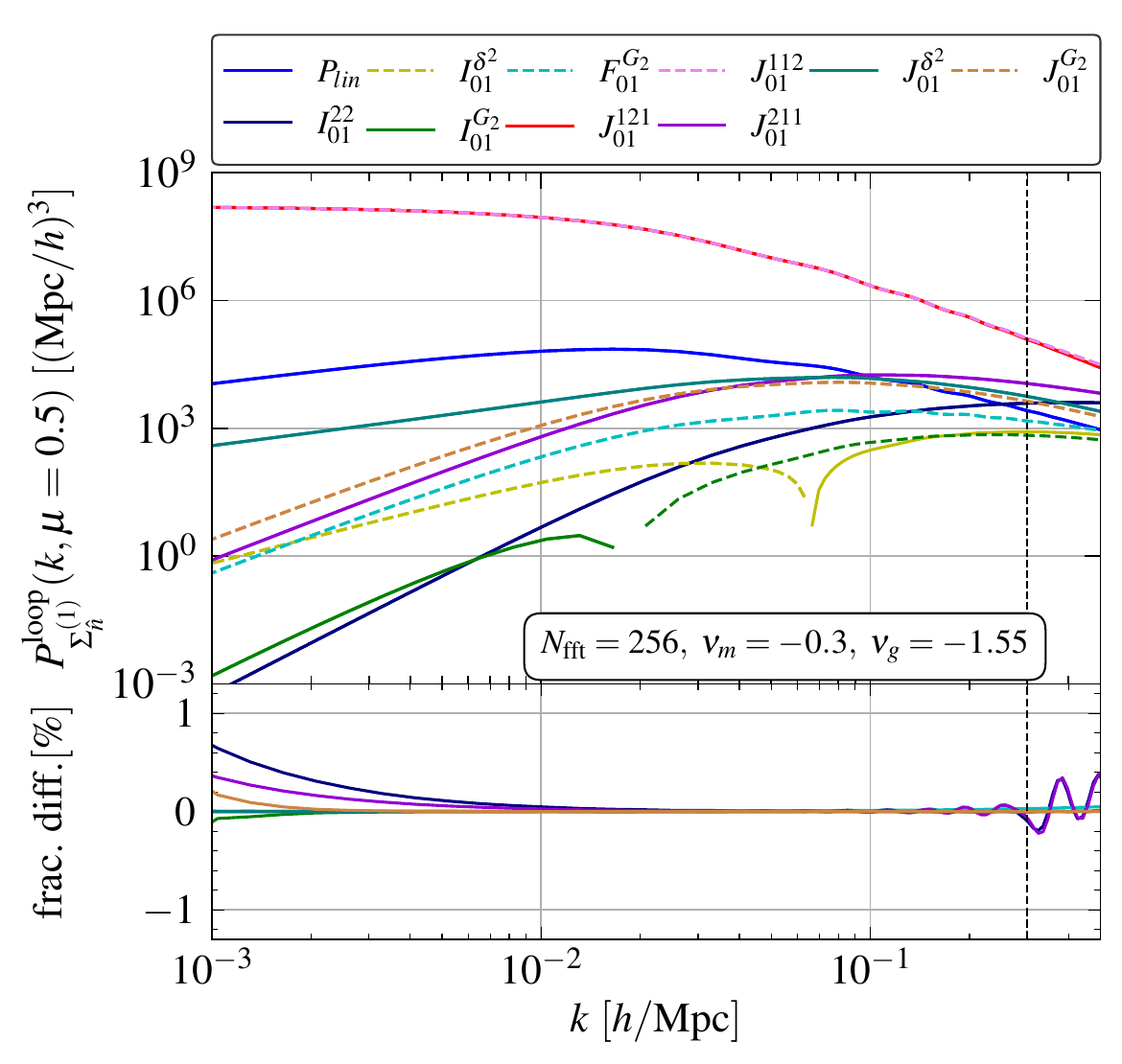}
    \includegraphics[width=0.496\textwidth]{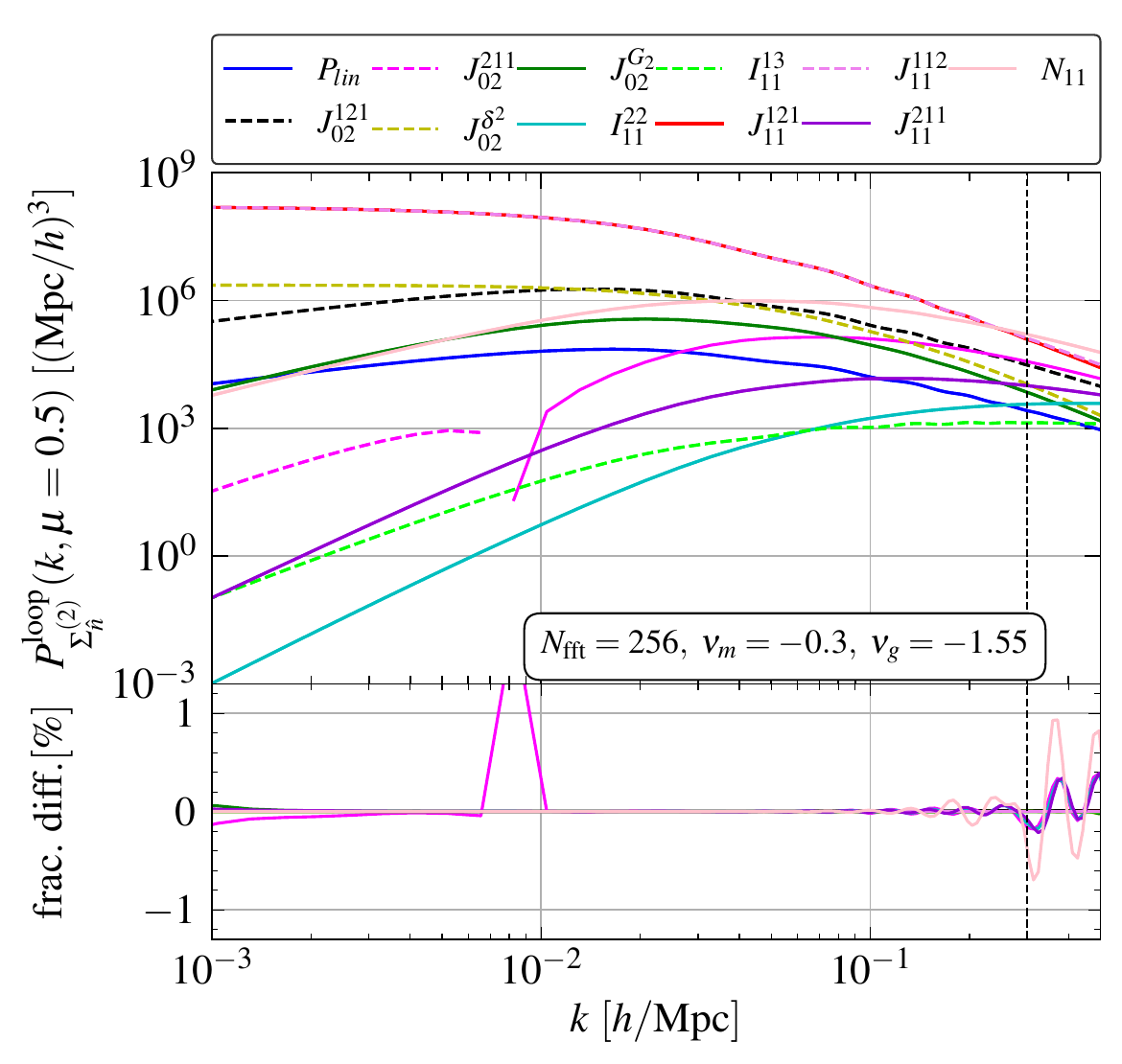}
    \caption{Comparison of the loop integrals contributing to the first (left) and second (right) velocity momenta defined respectively in eq. \eqref{eq:mom1} and \eqref{eq:mom2}, computed using either direct integration (DI) or the \texttt{FFTLog} approach. The top panels show individual contributions to the velocity moments. Dashed and solid lines correspond to negative and positive values, respectively. The vertical dashed line corresponds to $k=0.3 \ h/{\rm Mpc}$, which is the small-scale cutoff beyond which the perturbative one-loop model is expected to be inaccurate. The bottom panels show the fractional differences between DI and FFTLog results. On the scales most relevant for the modeling of nonlinear contributions ($0.01<k \ [{\rm Mpc}^{-1}h]<0.3$) the DI and FFTLog computations agree to better than 0.1\% (the spike on the residual difference plot on the right corresponds to a change of sign in the loop contributions and is not a point of concern).}
\label{fig:di_fft}\vspace{-0.15in}
\end{figure}

As an example of the validation of individual loop contributions, we show in figure \ref{fig:di_fft} various contributions to the first and second velocity momenta defined in \eqref{eq:mom1} and \eqref{eq:mom2}, computed either with the DI or \texttt{FFTLog} method, together with their residuals. We checked that all other moments agree at the same level. 

\subsubsection{Validation of {\sc CLASS-OneLoop} against {\sc CLASS-PT}}
\label{sec:code_validation}

As a benchmark for our implementation of one-loop EFT power spectrum in {\sc CLASS-OneLoop}, we compare the numerical evaluations of the real- and redshift-space galaxy power spectra against the publicly available code {\sc CLASS-PT}, which have many similarities and few technical differences that we now review.

First, both codes rely on the same Eulerian bias expansion, see eq. \eqref{eq:G_bias}.\footnote{We have also performed a comparison with {\sc Velocileptor} \cite{Chen:2020fxs} and found very good overall agreement. However, because of differences in the definition of bias parameters in {\sc Velocileptor} and {\sc CLASS-OneLoop} (or {\sc CLASS-PT}), we do not present this comparison in this section.}

Second, the modeling of the counter-terms and stochastic contributions slightly differs between {\sc CLASS-PT} and {\sc CLASS-OneLoop}. In the former, the counter-term parameters are defined for each Legendre multipole, while in our implementation they are associated to each velocity moment, see eq. \eqref{eq:expansion_ct}. After marginalization over these parameters, {\sc CLASS-PT} and {\sc CLASS-OneLoop} cover the same range of models, but a case-by-case comparison is difficult when the counter-terms are non-zero. Fortunately, such a comparison is not necessary for the sake of validation, because the counter-term parameters only multiply the linear power spectrum. Thus, for the purpose of the present comparison, we can fix all counter-terms to zero. We can actually do the same with the shot noise terms since they are not combined with any loop integral, see eq.~\eqref{eq:expansion_shot}. Without loss of generality, we set the values of the bias parameters to
\begin{align}
    b_1 = 1.8,\quad b_2 = -0.5,\quad b_{\mathcal{G}_2} = -0.05,\quad b_{\Gamma_3} = 0.08~.
\end{align}
These precise values are arbitrary, but ensure that the real- and redshift-space power spectrum are positive on all scales.

Third, the codes assume an Einstein-de Sitter (EdS) Universe to factorize the time and momentum dependence of the loops (replacing the EdS growth with the one in the true cosmology). Since the time-dependence is factorized out in a trivial way, it is sufficient to compare the two codes at a single redshift -- in this section, $z=0$.

Next, both {\sc CLASS-PT} and {\sc CLASS-OneLoop}  perform the loop calculations using the \texttt{\texttt{FFTLog}} algorithm introduced in ref. \cite{Simonovic:2017mhp}, with slightly different values of the FFT parameters. The goal of our comparison is to check that their predictions agree despite of these numerical differences. 

Finally, in both codes, the implementation of the IR-resummation is based on the wiggle no-wiggle split, with a suppression of the wiggle component, as described in Eqs. \eqref{eq:IR_real} and \eqref{eq:IR_redshift}. However, while {\sc CLASS-PT} uses a DST algorithm \cite{Hamann:2010pw, Baumann:2017gkg} to split the matter power spectrum into a wiggle and a no-wiggle component, our baseline implementation uses the Gfilter method \cite{Vlah:2015zda}.

Overall, we find an excellent agreement between the two codes, with small differences actually dominated by the use of different default algorithms for the wiggle/no-wiggle decomposition. Indeed, residual differences are even smaller when enforcing the same decomposition algorithm in both codes. Below, we describe in more detail the results of this comparison.

\begin{figure}[t]
    \centering
        \includegraphics[width=0.496\textwidth]{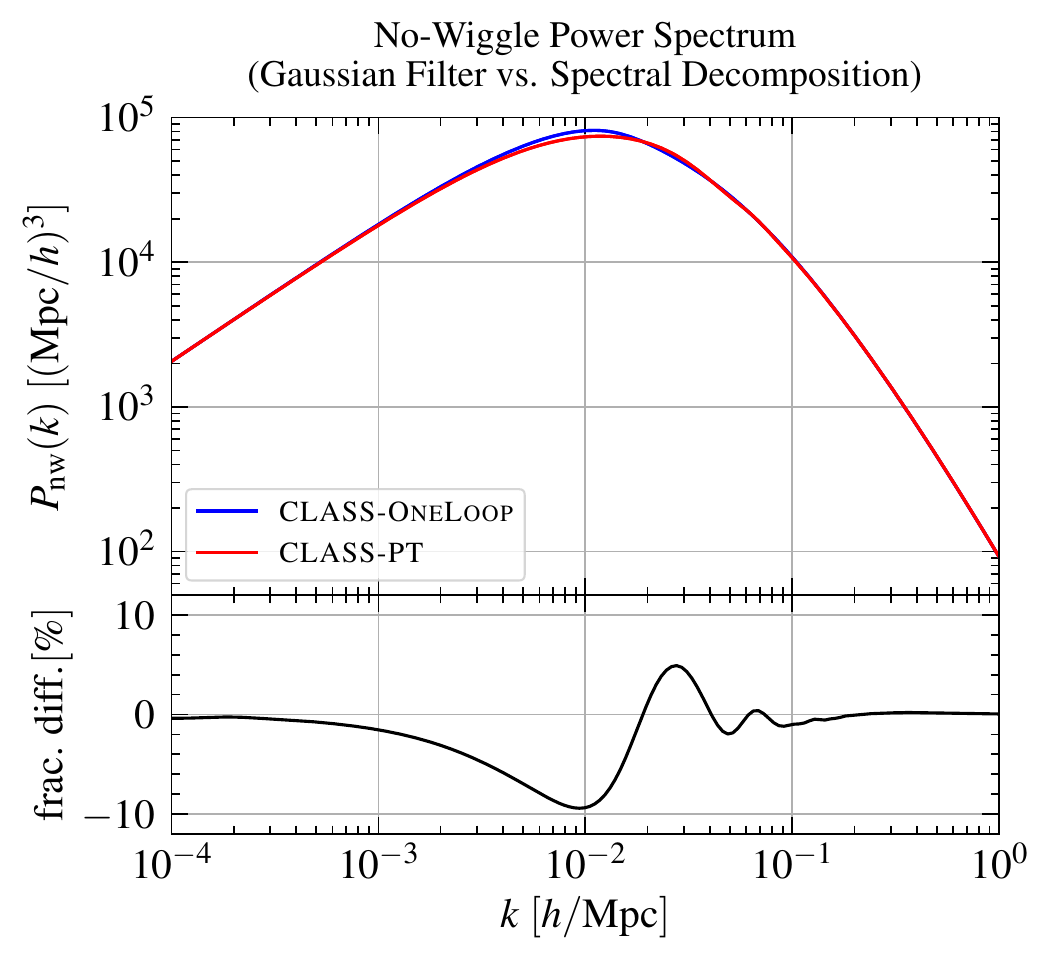}
        \includegraphics[width=0.496\textwidth]{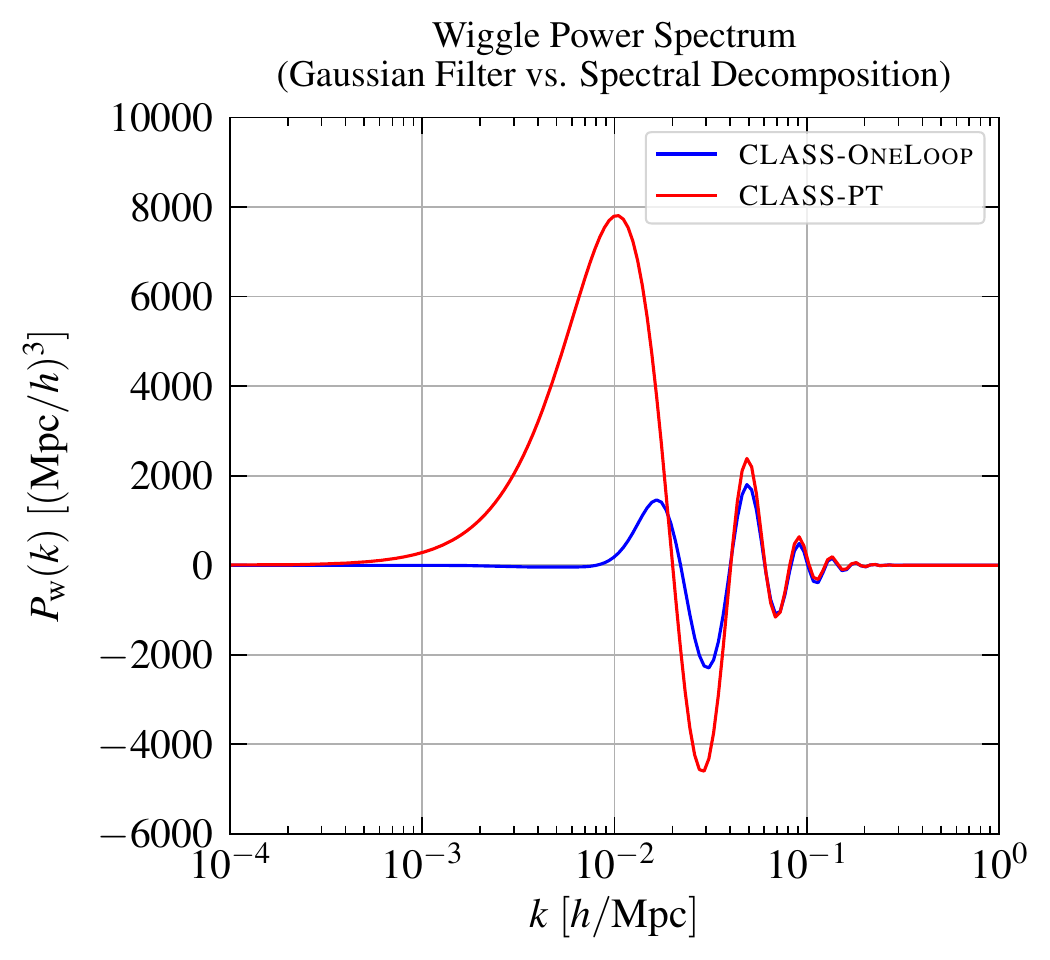}
    \caption{Comparison of the wiggle/no-wiggle decomposition of the linear matter power spectrum at redshift $z=0$ in {\sc CLASS-OneLoop} (blue) and {\sc CLASS-PT} (red). The left panel shows the broadband (no-wiggle) spectra and the right panel the extracted BAO features (wiggle spectra). For the broadband spectra, we also show the residuals (black). The two methods have a maximal 10\% discrepancy around the scale of the broadband peak, but for the purpose of fitting data the differences at smaller scales tend to be more relevant.}
    \label{fig:wnw_Comp_CLASS-PT}
\end{figure}

In figure \ref{fig:wnw_Comp_CLASS-PT}, we show the wiggle/no-wiggle decomposition of the linear matter power spectrum performed by default in {\sc CLASS-OneLoop} (red, Gfilter) and {\sc CLASS-PT} (blue, DST). The extracted broadband spectrum and BAO features are shown in the left and right plots, respectively. The fractional difference of the broadband is shown at the bottom panel of the left plot. We see that the DST artificially shifts the peak of the broadband spectrum towards smaller scales, resulting in an enhanced first BAO peak, shifted to larger scales compared to analytical predictions~\cite{Hu:1995en}. Furthermore, the amplitudes of the other BAO peaks differ in the two schemes, with higher amplitudes in the DST scheme.\footnote{See Appendix B of \cite{MoradinezhadDizgah:2020whw} for the comparison of these two algorithms with yet another approach based on fitting the broadband with a family of B-splines.} On linear scales, these discrepancies are not directly relevant, since after splitting the linear spectrum in a wiggle and no-wiggle term, one just adds them up again. Thus, the large discrepancies observed in figure \ref{fig:wnw_Comp_CLASS-PT} for the first oscillation(s) do not propagate to the IR-resummed one-loop power spectrum on the same scales. However, on smaller scales at which the wiggle part gets damped, small differences may propagate to the IR-resummed spectrum -- both at tree level and at the level of one-loop corrections, as suggested e.g., by eq. \eqref{eq:IR_redshift}. Thus, to evaluate the impact of discrepancies in the wiggle/no-wiggle decomposition, one needs to compare directly the one-loop spectra.

\begin{figure}[t]
    \centering
        \includegraphics[width=0.495\textwidth]{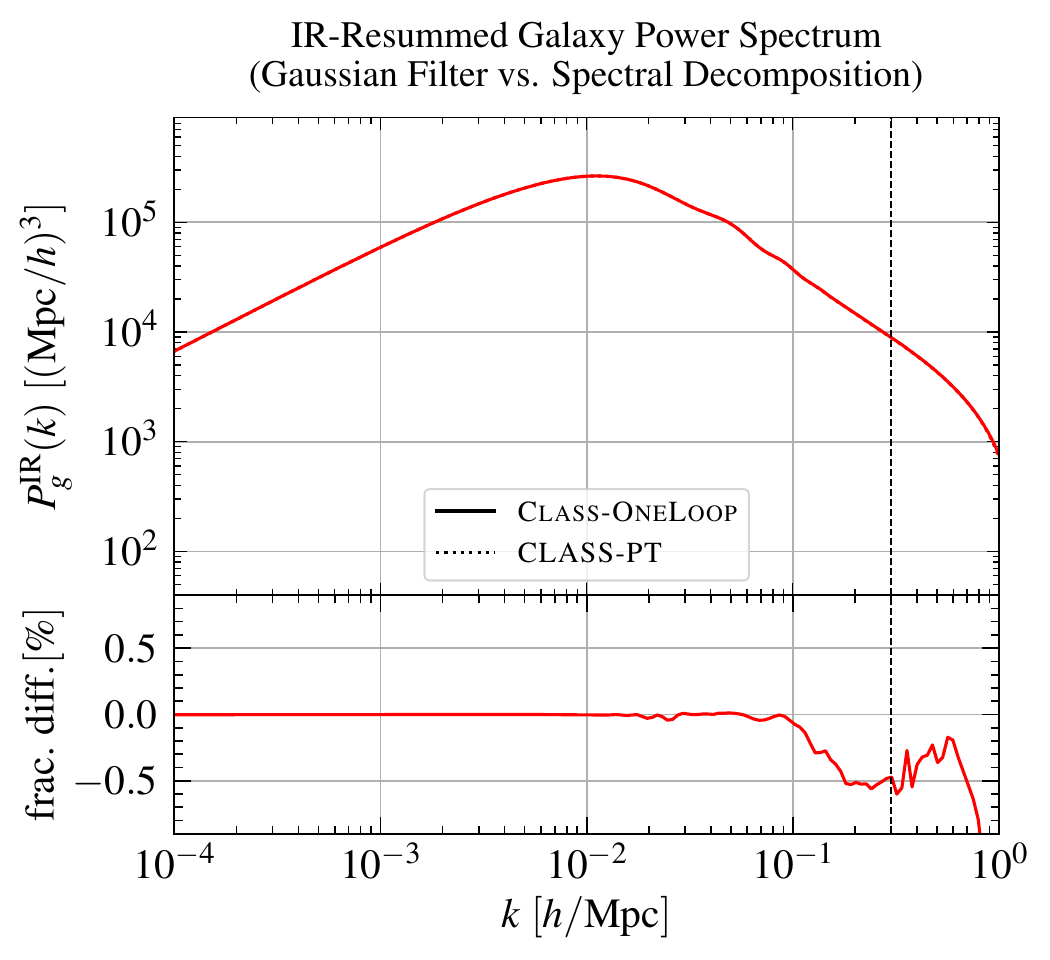}
        \includegraphics[width=0.496\textwidth]{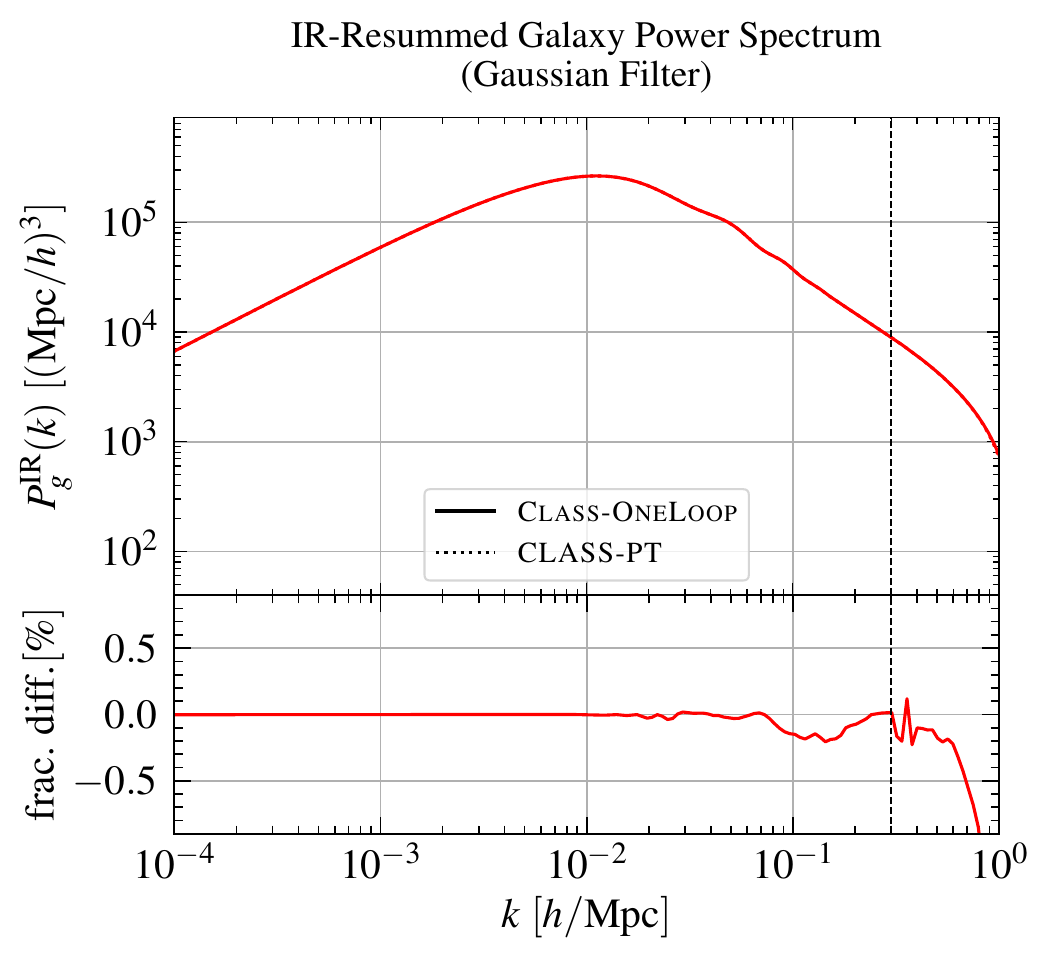}
        \includegraphics[width=0.496\textwidth]{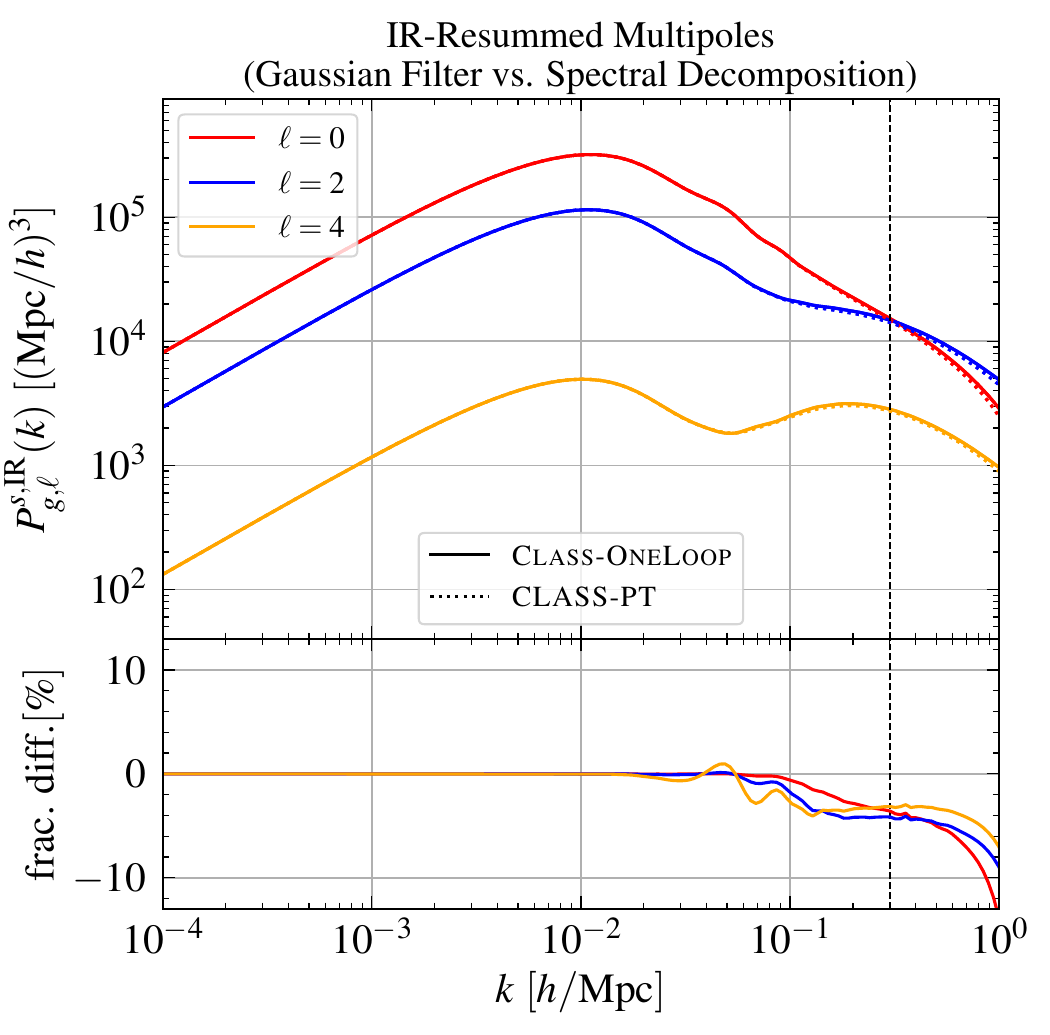}
        \includegraphics[width=0.496\textwidth]{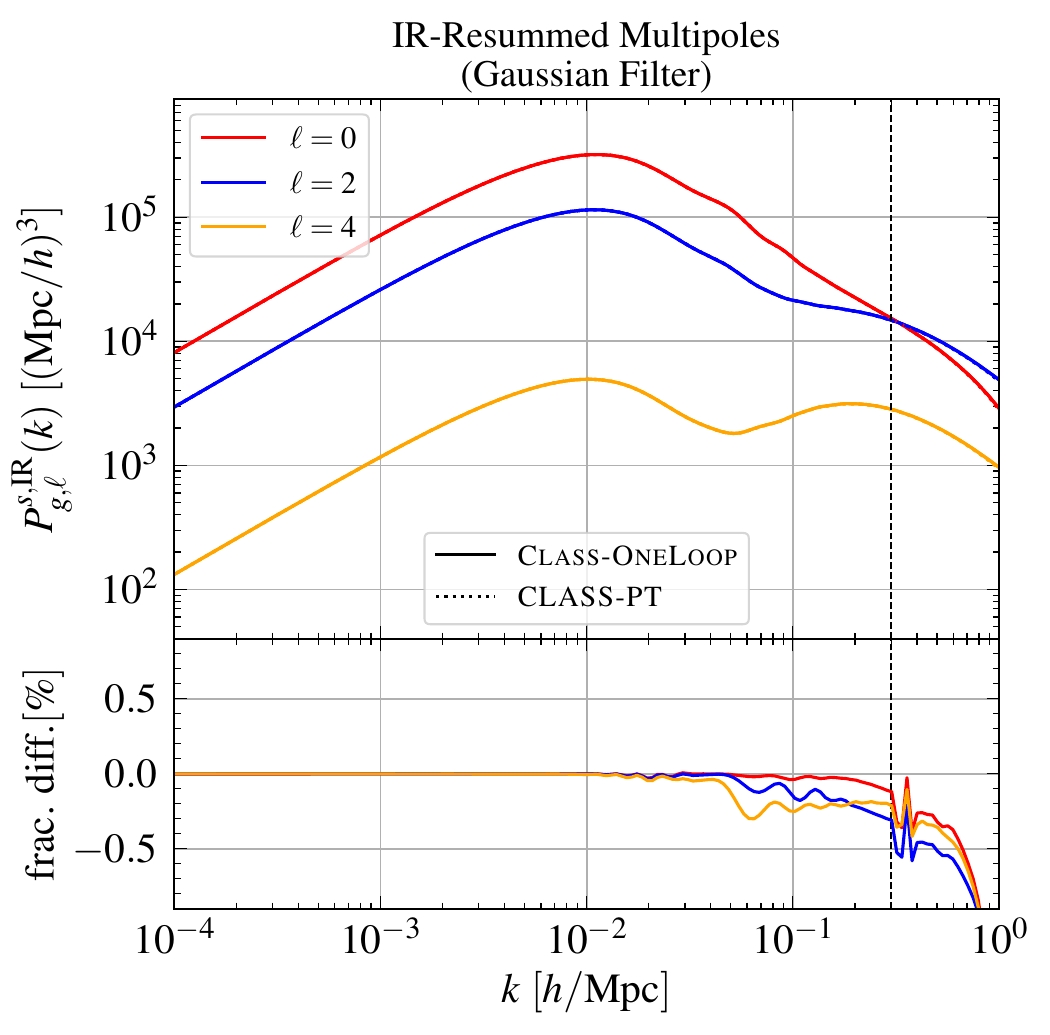}
    \caption{Comparison of the one-loop, IR-resummed galaxy power spectra computed by {\sc CLASS-PT} (dotted lines) and {\sc CLASS-OneLoop} (solid lines). The top row shows the real-space power spectrum, while the bottom row shows the first three power spectrum multipoles in redshift space in different colors. In the left panels, we stick to the default wiggle/no-wiggle algorithm of each code, while on the right we enforce the same Gfilter decomposition in both codes. In the upper panels of the top row, the {\sc CLASS-PT} and {\sc CLASS-OneLoop} predictions are indistinguishable by eye. The bottom panels show the fractional difference between the two codes. All the EFT counter terms and shot noise contributions are set to zero, while the biases are fixed to values described in the text. The vertical dashed line corresponds to $k=0.3 \ h/{\rm Mpc}$, which is the small-scale cutoff beyond which the perturbative one-loop model is expected to be inaccurate.}
    \label{fig:Galaxy_Comp_CLASS-PT}
\end{figure}

Figure \ref{fig:Galaxy_Comp_CLASS-PT} compares the output of the two codes at the level of the real-space power spectrum (upper plots) and redshift-space power spectrum expanded in Legendre multipoles $\ell=0$ (monopole), $\ell=2$ (quadrupole) and $\ell=4$ (hexadecapole). In the left plots, we stick to the default wiggle/no-wiggle decomposition algorithm, that is, Gfilter for {\sc CLASS-OneLoop} and DST for {\sc CLASS-PT}. In the right plots, we have replaced the wiggle/no-wiggle splitting of {\sc CLASS-PT} with the output of the split power spectrum from {\sc CLASS-OneLoop} obtained with the Gfilter method. 

In real space, the output of the two codes agree at the 0.1\%-level up to $k=0.1 \,h$/Mpc. Then, the difference increases to 0.5\% in the range $0.2\,h/\mathrm{Mpc} < k < 0.6\,h/\mathrm{Mpc}$. This discrepancy is however dominated by the wiggle/no-wiggle decomposition scheme since, after switching to the same Gfilter in both codes, we recover a remarkable 0.2\%-level agreement for any $k < 0.6\,h/\mathrm{Mpc}$. Note that the mismatch between the two codes in real space is mainly due to a slight difference in the predicted amplitudes of the one-loop contributions caused by the discrepancy in the extracted broadband (no-wiggle contribution), see the left panel of figure \ref{fig:wnw_Comp_CLASS-PT}. Fortunately, when using the default wiggle/no-wiggle algorithm of each codes, this discrepancy can, at least partially, be absorbed by the EFT counter term.

In redshift space, the discrepancy in the power spectrum multipoles is more significant when each code sticks to its default decomposition algorithm, with a relative difference of $\sim 5\%$ for $0.1\,h/\mathrm{Mpc} < k < 0.6 \,h/\mathrm{Mpc}$. Replacing the DST by a Gfilter in {\sc CLASS-PT}, we are back to 0.3\% level agreement between the two codes up to $k = 0.3 \,h/\mathrm{Mpc}$ and 0.6\% level agreement up to $k = 0.6 \,h/\mathrm{Mpc}$. Here again, like in real space, the discrepancy in the predicted amplitudes of the multipoles on small scales can be absorbed by the EFT counter terms. Apart from the difference in the predicted amplitude of the multipoles in the two codes, there is also a mild discrepancy in the BAO feature (most notable in the hexadecapole and to a lesser extent in the quadrupole) which partially remains even when using the same wiggle/no-wiggle algorithm. Two factors contribute to the mismatch on BAO scales; first, there is a slight difference in the amplitude of the higher BAO peaks, as shown in figure \ref{fig:wnw_Comp_CLASS-PT}, and second, the two codes make slightly different approximations in the implementation of eq. \eqref{eq:IR_redshift_approx}. The latter difference explains the residual discrepancy in the right lower panel of figure \ref{fig:Galaxy_Comp_CLASS-PT}.\footnote{In {\sc CLASS-PT}, instead of applying the suppression factor on the difference of the loops computed using the linear and no-wiggle matter power spectra (the curly bracket in the second line of eq. \eqref{eq:IR_redshift_approx}), the loops are directly computed using one insertion of wiggle and one insertion of no-wiggle power spectra, thus, discarding the $P_{\rm w}^2$ contributions \cite{Ivanov:2018gjr}.}

On the one hand, the above comparison confirms that the implementation of one-loop calculations -- including the {\tt FFTLog} expansion -- is very consistent across the two codes, up to a level of accuracy that is sufficient to fit the next generation of surveys. On the other hand, this comparison highlights that the predicted IR-resummed galaxy power spectra, particularly in redshift space, has a non-negligible dependence on the approach employed for the wiggle/no-wiggle decomposition, which must be defined with much care. At face value, one may argue that schemes yielding minimal features in the wiggle component (like our Gfilter) are expected to provide more accurate results than schemes enhancing these features (like the DST). Moreover, even if we consider that none of these two schemes is better than the other, the induced discrepancy would not necessarily play a role in the analysis of  next-generation surveys. Indeed, this difference is coming from the UV-sensitivity of the loop integrals and can therefore be absorbed by the corresponding counter-terms.

\subsubsection{Performance}
\label{sec:code_performance}

Table \ref{tab:speed} presents the wallclock time for computing the one-loop IR-resummed galaxy power spectrum in redshift space with $N_\mathrm{FFT}=128, 256$ or 512 Fourier components in the range  $k \in\left[ 10^{-5}, 51 \right] \,\mathrm{Mpc}^{-1}$ using the \texttt{FFTLog} method to compute the loop integrals. The quoted times include the full computation of FFTLog matrices and the calculation of each loop integral for an array of $k$ bins. It neglects the (very small) time for summing the loops and re-evaluating them for multiple $\mu$ values. If the loop matrices are cached between consecutive evaluations, the runtimes reduce drastically to those compiled in table~\ref{tab:speed_loops_only}. The reported timings were measured on the RWTH Cluster CLAIX-2018 subtracting the runtime of {\sc CLASS} for computing the linear matter power spectrum. The computing nodes contain 2 Intel Xeon Platinum 8160 Processors with 24 cores each and a total of 192GB of RAM. For these tests the code was compiled with {\sc icc} 2021.6.0 at optimization level 3.

As a reference, we also quote the timing for computing the model by performing direct numerical integration of the loops  using 4 {\sc OpenMP} threads. 

Multi-threading is very efficient for accelerating the evaluation of loop integrals as every mode is independent and requires the same amount of work, while the number of entries in each loop matrix vary between $\mathcal{O}(N_{\mathrm{FFT}})$ and $\mathcal{O}(N_{\mathrm{FFT}}^2)$.
\begin{table}[h]
    \centering
    \begin{tabular}{c|c|c|c||c}
         \hline
         & $N_{\mathrm{FFT}} = 128$ & $N_{\mathrm{FFT}} = 256$ & $N_{\mathrm{FFT}} = 512$ & Direct integration \\ \hline
        4 threads & $0.61 \pm 0.26$ & $2.05 \pm 0.15$ & $6.98 \pm 0.14$ & $\sim 600$ \\
        8 threads & $0.40 \pm 0.09$ & $1.38 \pm 0.12$ & $3.51 \pm 0.23$ & - \\
        16 threads & $0.52 \pm 0.11$ & $1.13 \pm 0.13$ & $2.00 \pm 0.22$ & - \\ \hline
    \end{tabular}
    \vspace{0.15in}
    \caption{Wallclock time in seconds for a single {\sc CLASS-OneLoop} run (deducting the linear part of the {\sc CLASS} workflow) for different numbers of Fourier components $N_\mathrm{FFT}$ and {\sc OpenMP} threads. Timings were measured on CLAIX-2018 systems averaging 150 consecutive runs.}
    \label{tab:speed}
     \vspace{-0.2in}
\end{table}

\begin{table}[h]
    \centering
    \begin{tabular}{c|c|c|c}
        \hline
         & $N_{\mathrm{FFT}} = 128$ & $N_{\mathrm{FFT}} = 256$ & $N_{\mathrm{FFT}} = 512$ \\ \hline
        4 threads & $0.101 \pm 0.008$ & $0.400 \pm 0.003$ & $1.467 \pm 0.085$ \\
        8 threads & $0.046 \pm 0.004$ & $0.212 \pm 0.018$ & $0.776 \pm 0.037$ \\
        16 threads & $0.028 \pm 0.003$ & $0.105 \pm 0.003$ & $0.382 \pm 0.0$ \\ \hline
    \end{tabular}\vspace{0.15in}
    \caption{Wallclock time in seconds for evaluating all loop integrals using the \texttt{FFTLog} method (deducting the linear part of the {\sc CLASS} workflow) on cached loop matrices for different numbers of Fourier components $N_{\mathrm{FFT}}$ and {\sc OpenMP} threads. Timings were measured on CLAIX-2018 systems averaging 150 consecutive runs}
    \label{tab:speed_loops_only}
\end{table}

The final version of {\sc CLASS-OneLoop}, to be publicly released, is expected to have further improvement in speed, and will be presented in more details in an upcoming publication \cite{release}. 

\section{Methodology and likelihoods} \label{sec:method_like}

In this section, we describe the relevant ingredients for the likelihood analyses, common to parameter inference from the power spectrum measured from N-body simulations and to MCMC forecasts for stage-IV surveys. Further details on each of these analyses will be presented in sections~\ref{sec:abacus} and \ref{sec:forecasts}.

\subsection{Likelihoods \label{sec:likelihoods}}

We assume a Gaussian form for the likelihood of the power spectrum wedges. Thus, the log-likelihood is given by 
\begin{equation}
    \ln \, \mathcal{L}  = -\frac{1}{2} \sum_{i,j,m,n} \Delta P_{im} C_{imjn}^{-1}  \Delta P_{jn} -\frac{1}{2} \operatorname{ln}\left( \frac{\operatorname{det} C}{\operatorname{det} C^{\mathrm{fid}}} \right),
\end{equation}
where $C_{ijmn}$ is the covariance matrix of the power spectrum wedges and $C^{\mathrm{fid}}_{ijmn}$ only serves as a normalization, and
\begin{equation}\label{eq:deltaP}
    \Delta P_{im} = P_{\rm data}(k_i,\mu_m) - P_{\rm theory}(k_i,\mu_m), 
\end{equation}
with $P_{\rm data}$ and $P_{\rm theory}$ being the ``measured" power spectrum wedges and their theoretical prediction, respectively. Since we neglect non-Gaussian contributions, the covariance matrix is diagonal. In MCMC forecasts, $P_{\rm data}$ refers to the fiducial spectrum generated using a theoretical model at a reference cosmology, while in the analysis of N-body simulations, it refers to the actual measurements of power spectrum wedges. In the latter case, to reduce the scatter of the measurements, we use the power spectrum wedges averaged over $N_R$ realizations $n$ of the simulation snapshots (see further details in section~\ref{sec:abacus}),
\be
\bar{P}_{\rm data}(k,\mu) = \frac{1}{N_R}\sum_{n=1}^{N_R} P_{\rm data}^{(n)}(k,\mu)~.
\ee
The measured power spectrum wedges are averaged over bins of widths of $\Delta k$ and $\Delta \mu$. To fit the power spectrum model to N-body simulations,  we also need to average the theoretical model over the same bins. Therefore, the theoretical power spectrum in eq. \eqref{eq:deltaP} is replaced by the bin-averaged quantity $\hat{P}_{\rm theory}$ computed from \cite{BOSS:2016teh}
\begin{equation}\label{eq:binned_P}
\hat{P}_{\rm theory}(k_i,\mu_m) = \frac{4\pi}{V_P(k_i)\Delta \mu} \int_{k_i - \Delta k/2}^{k_i + \Delta k/2}dk \ k^2\int_{\mu_m-\Delta \mu/2}^{\mu_m + \Delta \mu/2} d \mu \ P_{g,\rm AP}^{s,\rm IR}(k, \mu),
\end{equation}
where $V_P(k_i) = 4\pi k_i^2 \Delta k$ is the volume of each Fourier shell and $\Delta \mu$ is the width of each $\mu$ bin. We remind that $P_{g,\rm AP}^{s,\rm IR}$ are given in eq. \eqref{eq:Pkmu_AP}, and the integrals are evaluate over the wavenumbers and angles of the fiducial cosmology. We perform the bin averaging numerically using the $5$-point Gauss-Lobatto rule in each $\mu$ bin and the $8$-point Gauss-Legendre rule in each $k$ bin. 

The theoretical diagonal Gaussian covariance of the bin-averaged power spectrum wedges is given by \cite{Grieb:2015bia}
\begin{align}\label{eq:cov}
{\rm Cov}\left[\hat{P}_{\rm theory}(k_i,\mu_m), \hat{P}_{\rm theory}(k_j,\mu_n)\right] &= \frac{16 \pi k_f^3 \ \delta_{ij}\delta_{mn}}{V_P^2(k_i)(\Delta \mu)^2} \int_{k_i- \Delta k/2}^{k_i + \Delta k/2} \int_{\mu_m-\Delta \mu/2}^{\mu_m + \Delta \mu/2} d\mu dk \ k^2 \ \left[P_{g,\rm AP}^{s,\rm IR}(k,\mu)\right]^2.
\end{align}
where $\delta_{ij}$ and $\delta_{mn}$ are the Kronecker deltas. Since we account for the cosmology-dependence of the covariance in our analysis (i.e., we re-evaluate the covariance at each step in the MCMC chain), we account for the AP effect in the right side of eq. \eqref{eq:cov}. Therefore, $P_{g,\rm AP}^{s,\rm IR}$ and the integrals are evaluated at $k_{\rm fid}$ and $\mu_{\rm fid}$.

\subsection{Free parameters \label{sec:parameters}}

As described in section~\ref{sec:th}, in addition to the cosmological parameters, the one-loop EFT model of the power spectrum has a total of 12 nuisance parameters, consisting of four halo/galaxy biases, four stochastic parameters, and four EFT counter terms;
\begin{equation}
    \boldsymbol{\lambda}_{\rm bias} = \{b_1, b_2, b_{\cG_2}, b_{\Gamma_3}\}, \qquad \boldsymbol{\lambda}_{\rm stoch} = \{s_0, s_1, s_2, s_3 \}, \qquad  \boldsymbol{\lambda}_{\rm ct} = \{c_0, c_1, c_2, c_3\}.
\end{equation}
This large number of nuisance parameters may lead to parameter degeneracies, such that the effect of varying one parameter could be mimicked by a combined variation of other parameters. In this case, reducing the number of nuisance parameters can be beneficial in all respects: on the one hand, if done carefully, this reduction will not change the amount of information extracted from the data nor bias cosmological parameter inference, since the full parameter space is redundant; on the other hand, it will speed up the convergence of the MCMC algorithm, which can be strongly compromised by parameter degeneracies. As explained in the next section, we will perform a number of exploratory MCMC runs to establish which reduced parameter space ensures unbiased constraints on cosmological parameters, and ultimately vary 10 nuisance parameters for the main results presented in the paper.

The {\sc Abacus} simulations assume a $\Lambda$CDM cosmology, including massive neutrinos. To fit the power spectrum extracted from these simulations, we assume the same model varying five free parameters
\begin{equation}
\boldsymbol{\lambda}_{\rm cosmo} = \{\ln(10^{10} A_{\rm s}), n_{\rm s}, h, \omega_{\rm b}, \omega_{\rm cdm}\}, 
\end{equation}
with the notation $\omega_i \equiv \Omega_i h^2$. For forecasts of stage-IV galaxy surveys, we also assume a $\Lambda$CDM model as our fiducial cosmology. However, given the potential of stage-IV surveys to constrain DE, we fit the mock spectrum with an extended model allowing for dynamical dark energy. We consider a time-dependent equation of state, $w(a)$, obeying to the Chevallier–Polarski–Linder parameterization
\cite{Chevallier:2000qy,Linder:2002et} ($w(a)=w_0+w_a(1-a)$ with $a$ being the scale factor), and we vary 7 cosmological parameters
\begin{equation}
\boldsymbol{\lambda}_{\rm cosmo} = \{\ln(10^{10} A_{\rm s}), n_{\rm s}, h, \allowbreak \omega_{\rm b}, \omega_{\rm cdm}, w_0, w_a\}.
\end{equation}

\subsection{Priors \label{sec:priors}}

Our analysis of the
{\sc Abacus} simulations relies on splitting the full simulation box to small sub-volumes of $V_{\rm sub} = 1\ h^{-3}{\rm Gpc}^3$ (see section~\ref{sec:abacus}). This implies a relatively large power spectrum covariance, while limiting us to scales of $k_{\rm min} \geq 0.02 \ {\rm Mpc}^{-1}h$. Thus, in this case, we must impose a few informative priors on the cosmological parameters least constrained by the mock data in order to be able to infer other parameters. The question that we will ultimately address is: is our modelling of the galaxy power spectrum allowing us to infer these other parameters in an unbiased way?

\begin{table}[t]
    \centering
    \begin{tabular}{c|c|c}
        \hline
        Parameter & {\sc Abacus} fiducial & Prior \\
        \hline
        $\omega_{\rm b}$ & 0.02237 & $\mathcal N( 0.02237, 0.00036^2 )$ \\
        $\omega_{\rm cdm}$ & 0.12 & $\mathcal{U}( 0, 0.3 )$ \\
        $h$ & 0.6736 & $\mathcal{U}( 0.4, 0.8 )$ \\
        $\ln( 10^{10} A_{\rm s} )$ & 3.0364 & $\mathcal N( 3.044, 0.28^2 )$ \\
        $n_{\rm s}$ & 0.9649 & $\mathcal N( 0.9649, 0.084^2 )$ \\ \hline
    \end{tabular}\vspace{0.15in}
    \caption{Free cosmological parameters and their priors our fits to the {\sc Abacus} power spectrum. We assume a BBN prior on $\omega_{\rm b}$ and some loose {\it Planck} 2018 priors with a standard deviation multiplied by 20 on $\ln( 10^{10} A_{\rm s} )$ and $n_{\rm s}$. ${\cal N}(m,\sigma)$ denotes a normal (Gaussian) prior of mean $m$ and standard deviation $\sigma$ while ${\cal U}(a.b)$ denotes a uniform (flat) prior bounded to the range $[a,b]$.}
    \label{tab:priors}
    \vspace{-0.3in}
\end{table}

It is well-known that galaxy redshift surveys have limited sensitivity to the physical energy density of baryons, parameterized with $\omega_{\rm b}$. This density affects the shape of the  broadband power spectrum and the details of BAO peaks by a small amount, but as long as the error bars on the measured power spectrum are too large,  these effects cannot be separated from those of other parameters. Thus, when fitting the {\sc ABACUS} power spectrum, we impose an informative Gaussian prior $\omega_{\rm b}= 0.02233 \pm 0.00036$ (68\%CL) corresponding to the prediction of standard Big Bang Nucleosynthesis (BBN) \cite{Pisanti:2021}. 

Additionally, the limited volume and large covariance assumed in our fit to {\sc Abacus} makes it difficult to distinguish the effect of the primordial parameters $A_{\rm s}$ and $n_{\rm s}$ from that of bias parameters. In this case, we apply some very conservative priors on $\ln( 10^{10} A_{\rm s} )$ and $n_{\rm s}$ in the form of Gaussian distributions centered on the values measured by {\it Planck} 2018 but with a twenty times wider standard deviation. The goal of these priors is just to prevent the run from exploring a totally excluded region of parameter space.

Having assumed such priors, it will be interesting to check whether a fit of our model to the {\sc Abacus} power spectrum will return the corrected (fiducial) values of the remaining cosmological parameters $\omega_{\rm cdm}$ and $h$ for which we just assume top-hat priors in a very wide range, see table~\ref{tab:priors}.

In the case of our forecasts for a stage-IV galaxy survey, we only impose wide (uninformative) flat priors to check that the mock data allows to measure all cosmological parameters independently of the priors. 

We always impose flat uninformative priors on nuisance parameters in all our runs.\footnote{{\sc MontePython} has the option of running with uninformative flat prior, meaning: top-hat priors with boundaries far enough away from the preferred region to never be reached by the sampler. While some parameter inference engines such as  MultiNest require prior edges to be passed in an explicitly, this is not the case for the Metropolis-Hastings algorithm.}

We set the fiducial value of small-scale cut to $k_{\rm max} = 0.3\ {\rm Mpc}^{-1}h$. For {\sc Abacus} fits, we also run an additional \texttt{MCMC} chain with smaller value of $k_{\rm max} = 0.25\ {\rm Mpc}^{-1}h$ to test the dependence of the constraints on $k_{\rm max}$, while for the forecasts, we show the results for three additional values of $k_{\rm max}[{\rm Mpc}^{-1}h] = \{0.15, 0.2,0.25,0.3\}$. 

\section{Comparison with {\sc Abacus} N-body simulations}\label{sec:abacus}

Sections \ref{sec:code_validation_FFT_DI} and \ref{sec:code_validation} established the accuracy of theoretical predictions from {\sc CLASS-OneLoop} for the one-loop galaxy power spectrum by comparing against an external code. We will now validate the entire non-linear model and inference pipeline on a suite of N-body simulations. For this purpose, we use {\sc CLASS-OneLoop} along with the Bayesian inference package {\sc Monte Python} \href{https://github.com/brinckmann/montepython_public}{\faGithub}.\footnote{\url{https://github.com/brinckmann/montepython_public}}, using the Metropolis-Hasting algorithm to sample the likelihood. We constrain cosmological and nuisance parameters by fitting the halo power spectrum wedges measured on a suite of N-body simulations. To illustrate the importance of accurately modeling the loop contributions, we compare the parameter constraints obtained using the one-loop model against those derived from a simplified model that has often been used in previous forecasts for stage-IV surveys, and that we outline in appendix \ref{app:SPM}. 

\subsection{Simulation specifications \& power spectrum measurements}\label{sec:sims}

We make use of the halo catalogs from the {\sc AbacusSummit} simulation suite \cite{Maksimova:2021ynf} \href{https://github.com/abacusorg/abacussummit}{\faGithub},\footnote{\url{https://github.com/abacusorg/abacussummit}} obtained with the {\sc Abacus} N-body code \cite{Garrison:2021lfa}. More precisely, we 
use the \texttt{base} simulations, which evolve $(6912)^3$ particles with particle mass of $2 \times 10^9 \  h^{-1} M_\odot$ and a force softening at $7.2 h^{-1}$ proper kiloparsecs in a periodic cubic box of size $L_{\rm box} = 2 \ h^{-1}{\rm Gpc}$. The cosmology of the \texttt{base} simulations is set to the {\it Planck} 2018 best-fit $\Lambda$CDM model,\footnote{More specifically, the best-fit \texttt{base\_plikHM\_TTTEEE\_lowl\_lowE\_lensing} model is used.} which includes a single species of massive neutrinos with $\Sigma m_\nu$ = 0.06 eV. In the {\sc Abacus} code, neutrinos are modeled as a smooth component, including their effect on the Hubble expansion rate and neglecting their clustering. The initial conditions are set by scaling the linear matter power spectrum (combining the cold dark matter and baryon components) predicted by {\sc CLASS} at $z=1$ back to the initial redshift of $z = 99$. The back-scaling uses the linear growth function, including the suppression of growth induced by the smooth massive neutrino component. The initial conditions are generated using the public \texttt{Zeldovich-PLT} code \href{https://github.com/abacusorg/zeldovich-PLT}{\faGithub},\footnote{\url{https://github.com/abacusorg/zeldovich-PLT}} which includes first-order particle linear theory corrections \cite{Garrison:2018juw}. 

We focus on halo catalogs at redshift $z=0.5$ and use 25 realizations of \texttt{base} simulations (\texttt{AbacusSummit\_base\_c000\_ph\{001-0.25\}}), which rely on the same cosmology and only differ through their phase. We split each simulation box into 8 sub-boxes with a volume of $V_{\rm sub} = 1 \ h^{-3}{\rm Gpc}^3$.\footnote{We thank Chang-hoon Hahn for providing us the sub-volumes.} This enables us to have $N_R = 200$ realizations to reduce the scattering of the power spectrum measurements across individual simulations. The redshift-space halo power spectrum wedges are measured with fast Fourier transforms using the {\sc N-bodykit} package \cite{Hand:2017pqn} \href{https://github.com/bccp/N-bodykit}{\faGithub}.\footnote{\url{https://github.com/bccp/N-bodykit}} The halos are interpolated on a regular mesh of $256^3$ nodes using Triangular Shaped Cloud. Compensation is applied to deconvolve the effect of interpolation windows, and interlacing is used to reduce aliasing \cite{Sefusatti:2015aex}. The spectrum wedges are measured in 10 bins of $\mu$ and 64 bins of $k$, with width equal to twice the fundamental mode of the box, $\Delta k = 2 k_f = 2\pi/L_{\rm box}$. The redshift-space power spectrum is measured along a line-of-sight assumed to be the $y$ direction.

\subsection{Results: EFT and SPM fits of the power spectrum wedges} \label{sec:results}

\begin{figure}[t]
    \centering
    \includegraphics[width=0.98\textwidth]{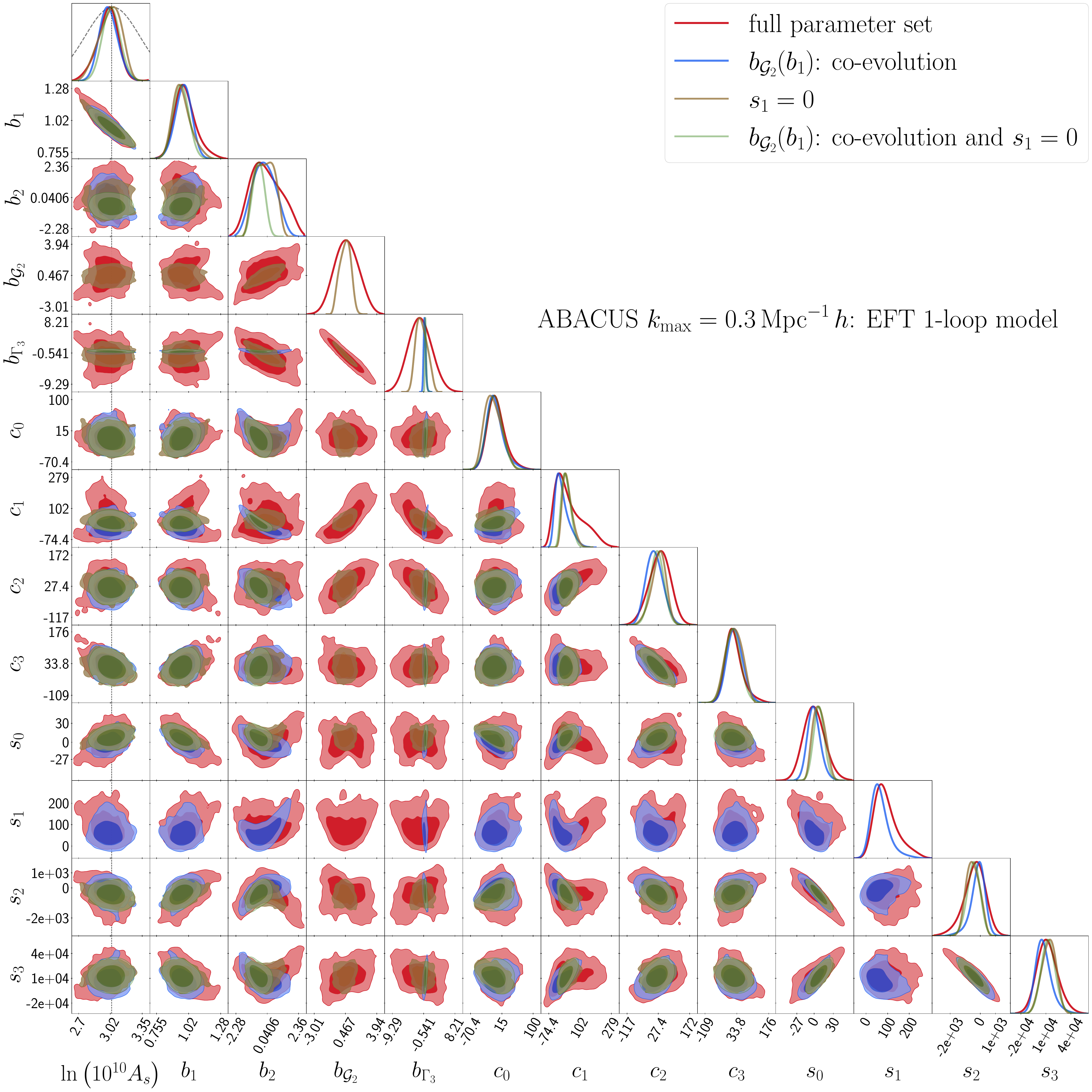}
    \caption{Marginalized 1- and 2-dimensional posterior distributions (68\% and 95\% confidence limits shown in darker and lighter colors) for $A_{\rm s}$ and EFT nuisance parameters inferred from the measured power spectrum wedges on {\sc Abacus} mocks at redshift of $z=0.5$. All other cosmological parameters are fixed to the fiducial values of the simulations. The small-scale cutoff is set to $k_{\mathrm{max}} = 0.3 \ h/{\rm Mpc}$. Different colors correspond to the results with different assumptions for the nuisance parameters; {\it red:} all nuisance parameters are varied independently, {\it blue:} the relation of $b_{\cG_2}(b_1)$ is set to co-evolution prediction, {\it gold:} the $k^2$ dependence of the shot noise contribution is set to zero ($s_1=0$), {\it green:} both assumptions are made.}
    \label{fig:nuisance_run}
\end{figure}

As briefly mentioned in section \ref{sec:method_like}, for the one-loop EFT model, given the relatively large number of nuisance parameters and the strong degeneracies among some of them, the first question to address is whether the size of the parameter space can be reduced without biasing the inferred cosmological parameters. We recall that the one-loop EFT model introduces 12 nuisance parameters per redshift bin. In the case of our {\sc Abacus} fits, we only consider a single redshift slice, but even with 12 independent nuisance parameters one may observe some strong degeneracies. Therefore, prior to doing an MCMC analysis varying all cosmological parameters, we perform a number of exploratory MCMC runs. To expedite the tests, we fix the four cosmological parameters ($\omega_{\rm b}$, $\omega_{\rm cdm}$, $h$, $n_{\rm s}$) and we focus on retrieving an unbiased constraint on $A_{\rm s}$ only. This strategy is numerically efficient because $A_{\rm s}$ is treated as a fast parameter within {\sc MontePython}, meaning that the linear power spectrum and the individual loop contributions only need to be computed a single time. Moreover, this test is sufficient to investigate all potential degeneracies and correlations among nuisance parameters.

As a matter of fact, we find that a full MCMC run with $A_{\rm s}$ plus 12 independent nuisance parameters is very expensive to converge due to severe degeneracies (especially between the bias parameters $b_{\cG_2}$ and $b_{\Gamma_3}$ and to a lesser extent between $b_{\cG_2}$ and $b_2$) and the strongly non-Gaussian contours (especially those involving the shot noise parameter $s_1$), as can be checked in figure \ref{fig:nuisance_run}.

To reduce the dimensionality of the nuisance parameter space, one has the option of fixing some of them or assuming relations among them according to well-motivated theoretical priors. Among the nuisance parameters of the power spectrum model, the second and third-order tidal biases, $b_{\cG_2}$ and $b_{\Gamma_3}$, are known to be highly degenerate, rendering the MCMC analysis very challenging. A theoretically motivated assumption is to relate both or one of these biases to linear bias $b_1$ using the so-called co-evolution model. This model underscores that even in the absence of initial tidal biases in Lagrangian space during formation, the late-time description of the galaxy/halo density field through local-in-matter biases $b_n$ is altered by gravitational evolution, introducing tidal contributions in the bias expansion in Eulerian space. In this case, the conservation of the total number of objects implies that the late-time non-local bias parameters are all related to the linear bias $b_1$ through
\be
b_{\cG_2} = -\frac{2}{7}(b_1-1), \qquad b_{\Gamma_3} = -\frac{1}{6}(b_1-1) -\frac{5}{2} b_{\cG_2}~.
\ee
These relations have been tested against simulations e.g. in ref.~\cite{Lazeyras:2017hxw}. The measured biases in halo catalogs constructed from N-body simulations were found to be in good agreement with the above relations, especially the one between $b_{\cG_2}$ and $b_1$. We thus repeated our MCMC run with $b_{\cG_2}$ given as a function of $b_1$ by the co-evolution model rather than being varied independently. The results are shown in blue in figure \ref{fig:nuisance_run}.  In this case, as expected, the main parameter degeneracies disappear and we obtain definite predictions for $b_{\Gamma_3}$ and $b_2$, while the posterior of the cosmological parameter $A_{\rm s}$ and even for the bias $b_1$ remains stable. Utilizing the co-evolution prediction for $b_{\Gamma_3}$ instead leads to a slight shift of roughly $0.4\, \sigma$ of the mean of the inferred $A_{\rm s}$-distribution. We conclude that for {\sc Abacus} simulations using the relation $b_{\cG_2}(b_1)$ allows for a more efficient fit without altering the information extracted from the data. This conclusion may not hold for other simulated and observational data sets and should be verified accordingly.

We now discuss the role of the nuisance parameters describing stochastic corrections. Eq. \eqref{eq:expansion_shot} provides the first terms in a Taylor expansion of the function $P_{\rm shot}(k,\mu)$ in powers of $\mu^2$ and $k^2$. At the one-loop order, it is clear that we should have stochastic corrections of order $\mu^0$, accounting for shot noise in the density field, of order $\mu^2$, accounting for shot noise in the velocity field, and of order $\mu^4$, accounting for corrections to 4th-order velocity moments. In an expansion with respect to $k^2$, the $s_0$ term is the leading-order term in $\mu^0$; the $s_2$ term is the leading term in $\mu^2$; and the $s_3$ term is the leading term in $\mu^4$. Thus, it is clear that these three terms play an important role in the one-loop galaxy spectrum model. Then, at each order in the $\mu^2$ expansion, there can be higher-order corrections in the $k^2$ expansion. For instance, the term containing $s_1$ is the next-to-leading $\mu$-independent term in an expansion in powers of $k^2$. It is thus not obvious that this term plays a significant role in the model. There are even higher-order terms in $\mu^0$, $\mu^2$ and $\mu^4$ that we did not even include in the shot noise expansion of eq.  \eqref{eq:expansion_shot}. The question is whether considering a non-zero $s_1$ is really necessary in order to accurately model the one-loop spectrum. 

The two runs mentioned above included a marginalisation over $s_1$. They actually show that there is no significant correlation between $s_1$ and $A_{\rm s}$. Additionally, they reveal subtle degeneracies between the stochastic parameter $s_1$ and several other parameters (most notably $b_2$, $b_{\cG_2}$, $b_{\Gamma_3}$ and $c_1$), leading to strongly non-elliptical 2D contours (see figure~\ref{fig:nuisance_run}). This is a sign that $s_1$ is degenerate with other parameters combinations and contains redundant information. To check this, we performed a run with $s_1$ set to zero (but still an independent  bias $b_{\cG_2}$).
The result, shown in gold color in figure \ref{fig:nuisance_run}, confirms that the $A_{\rm s}$ posterior is unaffected by this reduction of the nuisance parameter space. Finally, we run again with the two priors  $b_{\cG_2}(b_1)$ and $s_1=0$, that is, with 10 nuisance parameters. This leads to the green contours of figure \ref{fig:nuisance_run}, which are now all nearly elliptic. In this case, the convergence of the MCMC run is very fast compared to the case with 12 nuisance parameters, while the posterior of $A_{\rm s}$ is unaffected. We will thus adopt such a reduction of nuisance parameter space in the next runs.

\begin{figure}[t]
    \centering
    \includegraphics[width=0.496\textwidth]{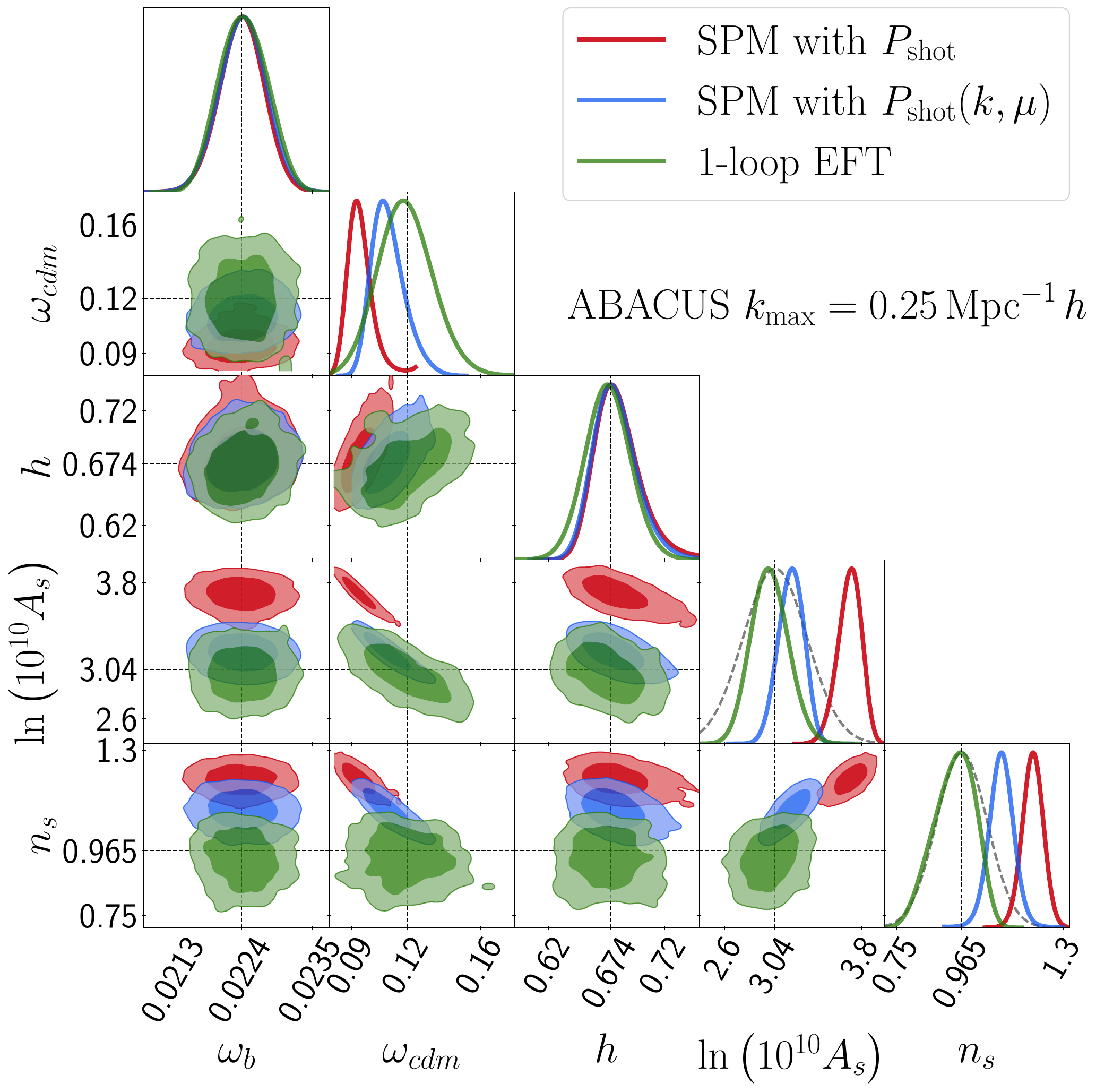}
    \includegraphics[width=0.496\textwidth]{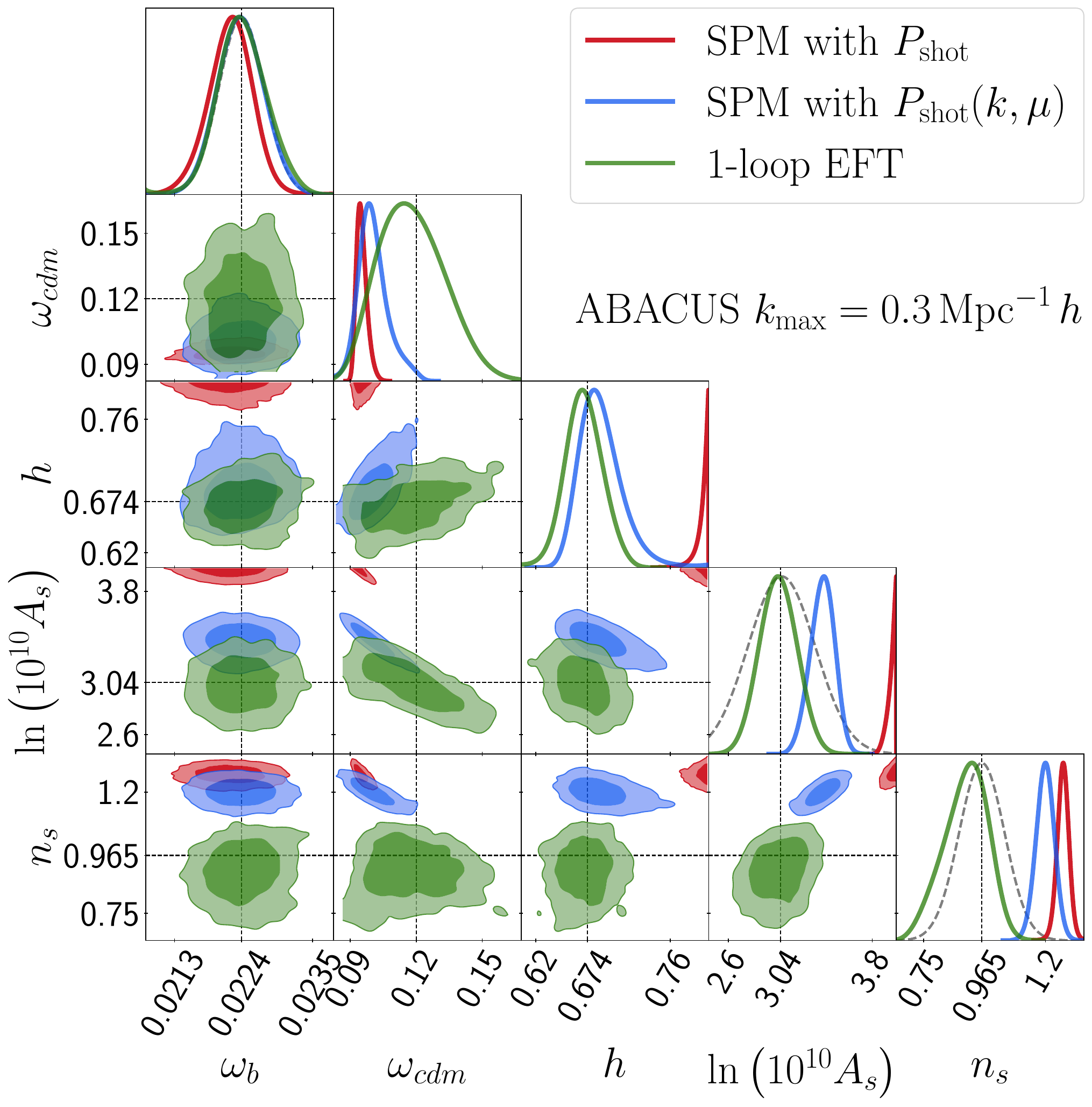}
    \caption{Marginalized 1- and 2-dimensional posterior distributions of cosmological parameters inferred from the measured power spectrum wedges on the {\sc Abacus} mocks at redshift of $z=0.5$. The green colour refers to a fit of the one-loop EFT model to the mock data. The red and blue contours show the results derived from the SPM described in appendix \ref{app:SPM}, with either a single shot noise parameter $s_0$ (red) or a set of stochastic terms $\{s_0, s_2, s_3\}$ (blue). The small-scale cutoff is set to $k_{\rm max} = 0.25 \, h \, \mathrm{Mpc}^{-1}$ on the left and to $k_{\rm max} = 0.3 \, h \, \mathrm{Mpc}^{-1}$ on the right. The Gaussian priors of table \ref{tab:priors} on $\omega_{\rm b}$, $n_{\rm s}$ and $A_{\rm s}$ are indicated in grey. The constraint on $\omega_{\rm b}$ is dominated by the BBN prior (such that the $\omega_{\rm b}$ posterior curve is hidden behind the prior curve). For other parameters, we adopt flat, wide and uninformative priors.}
\label{fig:N-bodyfit_abacus_cosmo}
\end{figure}

We then proceed to MCMC runs with the full set of cosmological parameters and priors listed in table \ref{tab:priors}. The runs reach stable posterior distributions with a convergence statistic of $|R - 1| < 0.05$ after about $10^6$ accepted points per chain.\footnote{The oversampling factor of the nuisance parameters w.r.t. the cosmology was set to $12$.} The results are displayed in green color in figure \ref{fig:N-bodyfit_abacus_cosmo} with two choices for the cutoff in wavenumber space: a more conservative cut at $k_{\rm max}=0.25 \ h/{\rm Mpc}$ (left panel) and a more aggressive cut at $k_{\rm max}=0.3 \ h/{\rm Mpc}$ (right panel). For clarity, we only include cosmological parameters in these plots. The full triangle plot including cosmological and nuisance parameters is shown in figure~\ref{fig:abacus-1loop-full} of appendix \ref{app:nuisance}. The most striking conclusion is that the fiducial values of cosmological parameters used in the simulation (black dotted lines) are always perfectly recovered in the two cases. This is a clear sign of success for the one-loop EFT model, which correctly captures the non-linear effects present in the simulation.  Due to the restricted volume of the simulation data, the inference of $\omega_{\rm b}$ and $n_{\rm s}$ is dominated by the priors. Instead, some information regarding $A_{\rm s}$ is extracted from the data since the posterior is narrower than the prior. Finally, the correct value of $\omega_{\rm cdm}$ and $h$ is inferred entirely from the data.

We finally compare the result of parameter inference based on the one-loop EFT model to the one of the simplified phenomenological model (SPM) described in appendix~\ref{app:SPM}, which has been adopted in several previous forecasts (e.g. \cite{Euclid:2019clj,Euclid:2023pxu}).
In this case, the model parameters that are marginalised over are just the linear bias $b_1$, the velocity dispersion $\sigma_{\rm p}^2$ and the shot noise parameter $s_0$. In figure \ref{fig:N-bodyfit_abacus_cosmo}, the red colour shows the result obtained when applying the SPM model literally, with the single stochastic parameter $s_0$ accounting for shot noise in the density field.  The blue colour refers to an ``enhanced SPM'' in which we adopt the same set of stochastic terms depending on $\{ s_0, s_2, s_3\}$  as in the one-loop EFT model in order to obtain more conservative results. We also tried to marginalise over the pairwise velocity dispersion $\sigma_{\rm v}^2$, instead of fixing it to the value predicted in Eq.~\eqref{eq:Sigma_integral}, but this did not improve the results as we describe below. 

With $k_{\rm max}=0.25 \ h/{\rm Mpc}$, the failure of the SPM model is already very striking: the posteriors and contours are very narrow and suggest totally incorrect values of cosmological parameters. With the minimal shot noise model, the SPM prediction is $4.1\sigma$ away from the true value for the parameter $\omega_{\rm cdm}$. The posterior of $A_{\rm s}$ and $n_{\rm s}$ are also strongly biased despite of the correctly-centered priors. Adding an additional set of stochastic parameters helps to partially decrease the bias of the SPM posteriors, although the $n_{\rm s}$ posteriors stays more than 2$\sigma$ away from the true values. We may also evaluate the goodness-of-fit of each model using a $\chi^2$-statistics computed at maximum likelihood using the analytical (diagonal) covariance matrix. The minimum $\chi^2$ value in each case is reported in Table~\ref{tab:chi2_bestfit}. With $k_{\rm max}=0.25 \,h/{\rm Mpc}$, the one-loop EFT model is able to achieve a minimum $\chi^2$ of $5.1$, to be compared with a total of $180$ data points. Instead, the SPM model with a minimal shot noise model cannot achieve better than $\chi^2=317$, or $39.2$ with the enhanced shot noise prescription. We would therefore still consider the EFT model to be successful if the same measurements were been taken in a survey volume that is $35$ times larger than the {\sc Abacus} simulation sub-boxes employed here. The SPM on the other hand is already providing a bad fit at a survey volume of $1\, \mathrm{Gpc}^3\, h^{-3}$. 

With $k_{\rm max}=0.30 \, h/{\rm Mpc}$, the SPM model performs even worse, because it can never account for the shape of the non-linear power spectrum at the smallest scales. The chains are driven to the largest values of $h$ compatible with our top-hat prior range ($h<0.8$), while other cosmological parameters also drift to extreme values. Table~\ref{tab:chi2_bestfit} shows that $\chi^2$ values cannot decrease below 1036 (resp. 147) in the minimal (resp. extended) shot noise model, while the one-loop model achieves $\chi^2=7.44$ for 220 data points. We checked that the situation does not improve when floating one additional parameters $\sigma_{\rm v}^2$ in the SPM model. In this case, the MCMC results are even worse, with $\sigma_{\rm v}^2$ and some cosmological parameters drifting very far from their expected value. This shows that there is no straightforward way to cure or absorb the incorrect modelling performed in the SPM case, and that this model should not be used to fit real data.

\begin{table}[h]
    \centering
    \begin{tabular}{c|c|c}
        Model & $\chi^2$ for $k_{\mathrm{max}} = 0.25 \, \mathrm{Mpc}^{-1}\, h$ & $\chi^2$ for $k_{\mathrm{max}} = 0.3\, \mathrm{Mpc}^{-1}\, h$ \\ \hline
        SPM with $P_{\mathrm{shot}}$ & 316.9 & 1036.7 \\
        SPM with $P_{\mathrm{shot}}(k, \mu)$ & 39.2 & 147.6 \\
        one-loop EFT & 5.1 & 7.4 \\ \hline
        data points & 180 & 220 \\
    \end{tabular}\vspace{0.1in}
    \caption{$\chi^2$-values for the ABACUS fits evaluated at the best-fitting model within MCMC chains.}
    \label{tab:chi2_bestfit}
\end{table}

\begin{figure}[t]
    \centering
    \includegraphics[scale=0.6]{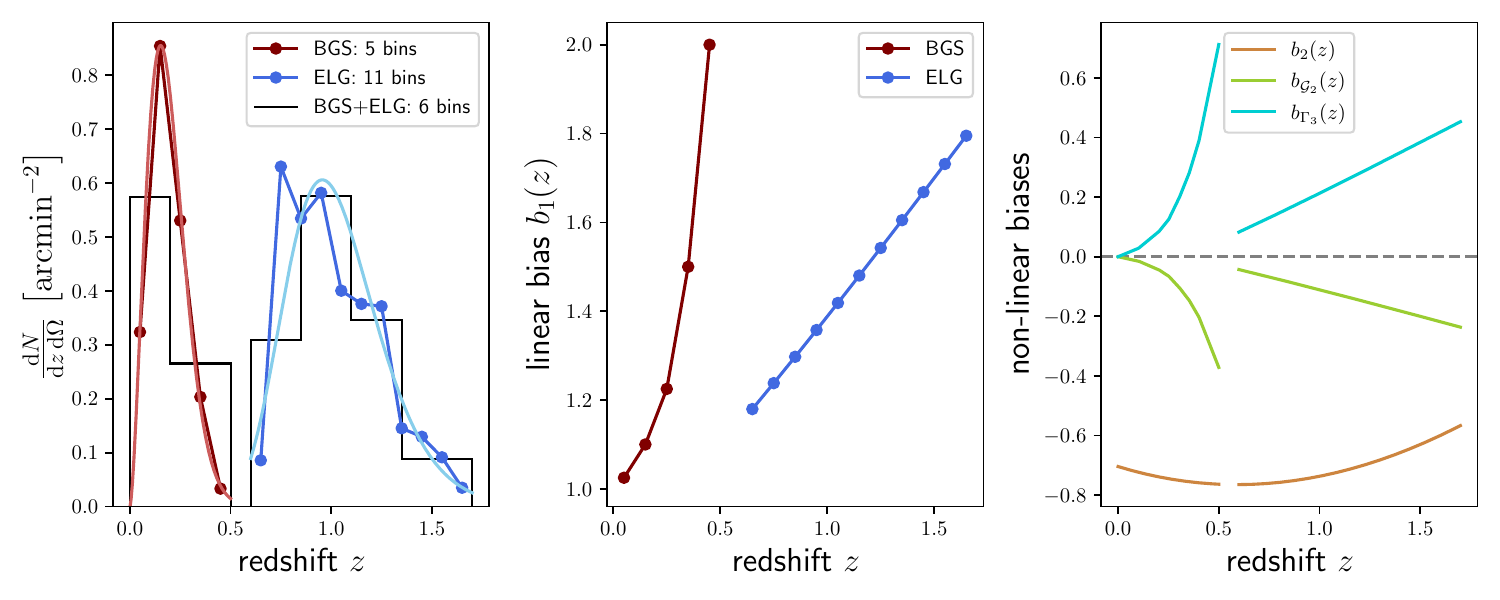}\vspace{-0.1in}
    \caption{Expected redshift distribution, linear biases, and non-linear biases in co-evolution for the low-redshift BGS ($0 < z < 0.5$) and the high-redshift ELG sample ($0.6 < z < 1.6$) in the two different binning schemes. The colored point markers show values for the original thin DESI bins of $\Delta z = 0.1$ while the black histograms show the 6 wider bins with $0.2 < \Delta z < 0.35$. The light colored lines represent the analytical fit given in eq. \eqref{eq:dNdz_fit}.}
    \label{fig:galaxy_number_rebinning}\vspace{0.1in}
\end{figure}

\section{\texttt{MCMC} forecast for stage-IV galaxy surveys} \label{sec:forecasts}

Next, we use {\sc CLASS-OneLoop} together with the Monte Carlo code {\sc MontePython} to perform an \texttt{MCMC} forecast for stage-IV galaxy surveys, considering a survey with the specifications matching those of DESI \cite{DESI:2016fyo}. We assume a Gaussian likelihood for the power spectrum wedges and perform a forecast using the EFT one-loop model described in section \ref{sec:th}. Additionally, we augment the model to account for the redshift uncertainties, which smear the galaxy density field along each line of sight, as an overall exponential suppression, $F_z(k, \mu) = {\rm exp}\left(-k^2\mu^2\sigma_{r}^2\right)$ that multiplies the one-loop power spectrum. Here, $\sigma_{r}^2 = c(1+z)\sigma_{0,z}/H(z)$ and $\sigma_{0,z}$ is the standard deviation of expected redshift errors, set to $\sigma_{0,z} = 0.001$ for a DESI-like survey. The Markov chain sampling is performed using the Metropolis-Hastings algorithm on fiducial data generated with the same model. 

\subsection{Survey specifications and forecast assumptions}\label{sec:specs_assump}

DESI is a five-year ground-based survey, which completed its five-month Survey Validation in May 2021 \cite{DESI:2023ytc}, thus, it is the first stage-IV galaxy survey to become operational. The DESI instrument is mounted on U. Mayall 4-meter Telescope at Kitt Peak National Observatory's Nicholas in southern Arizona and will obtain optical spectra for stars, galaxies, and quasars over approximately 14,000 square degrees of the sky corresponding to a sky fraction of $f_{\mathrm{sky}} = 0.34$. The extragalactic spectroscopic sample consists of four distinct classes of extragalactic sources; bright galaxy sample (BGS), luminous red galaxies (LRGs), star-forming emission line galaxies (ELGs), and quasi-stellar objects (QSOs). The samples span the redshift-ranges of $0.05<z<0.4$, $0.4<z<1$, $0.6<z<1.6$, and $0.9<z<2.1$, respectively \cite{DESI:2016fyo}.

In the forecasts presented below, we consider BGS and ELG samples, individually and combined. Given that the two samples do not have redshift overlap we treat them as uncorrelated. We use the original specifications of ref. \cite{DESI:2016fyo} for the redshift ranges, the redshift distributions $N(z)$, the linear biases for each redshift bin, and the redshift error. Reference \cite{DESI:2016fyo} assumes 5 bins for the BGs and 11 bins for the ELGs, each with a width of $\Delta z = 0.1$. In our analysis, in order to lower the computational burden of varying all 10 nuisance parameters in each redshift bin, we redistribute the samples over a smaller number of bins. We consider 2 bins for the BGs and 4 bins for the ELGs with a redshift width between $0.25$ and $0.35$.\footnote{Bins at the edges of the redshift range were chosen slightly wider in order to smoothen the galaxy number distribution.} To obtain the correct galaxy distribution in the new bins, we fit the number of galaxies per redshift slice, provided in ref. \cite{DESI:2016fyo}, with the model function which is a generalization of the form used in ref. \cite{Euclid:2019clj},
\be\label{eq:dNdz_fit}
\frac{\mathrm{d}N_{\mathrm{model}}}{\mathrm{d}z\, \mathrm{d}\Omega} = a z^b e^{-c z^d},
\ee
and integrate over the bin width and angle to obtain the total number of galaxies in the corresponding bin. The amplitude parameter $a$ was rescaled to keep the total number of galaxies in each sample consistent with the original specifications. The fitted redshift distribution is shown in figure \ref{fig:galaxy_number_rebinning}, and the new binning specifications are tabulated in detail in tab. \ref{tab:forecast_rebinning}.

\begin{table}[]
    \centering
    \begin{tabular}{cc|c|c|c}
        \hline
        $z_{\mathrm{min}}$ & $z_{\mathrm{max}}$ & $V_c(z)$ $[\mathrm{Gpc}^3] \, h^{-3}$ & $n_g(z)$ $[10^{-4}\, \mathrm{Mpc}^{-3} \, h^3]$ & $b_1(z)$  \\ \hline
        $0.0$ & $0.2$ & $0.2633$ & $219.831$ & $1.05$ \\
        $0.2$ & $0.5$ & $2.9540$ & $13.582$ & $1.50$ \\ \hdashline
        $0.6$ & $0.85$ & $6.680$ & $5.811$ & $1.223$ \\
        $0.85$ & $1.1$ & $9.045$ & $8.010$ & $1.373$ \\
        $1.1$ & $1.35$ & $10.811$ & $4.023$ & $1.527$ \\
        $1.35$ & $1.7$ & $17.081$ & $0.915$ & $1.716$ \\ \hline
    \end{tabular}\vspace{0.15in}
    \caption{DESI-like 6 bin survey specifications obtained from analytical fits to the differential galaxy number and averaging it inside of these wide bins. The total number of galaxies was kept constant during the re-binning procedure at $9.8 \times 10^{6}$ and $17.04 \times 10^{6}$ for the BGS and ELG samples respectively. $V_c(z)$ is the comoving volume of the bin and $b_1(z)$ the linear bias which was linearly interpolated to the bin center from the original DESI specifications.} \vspace{-0.15in}
    \label{tab:forecast_rebinning}
\end{table}

\begin{table}[]
    \centering
    \begin{tabular}{c|cccc}
        \hline
        Sample & $a$ & $b$ & $c$ & $d$ \\ \hline
        BGS & $1.444 \cdot 10^{5}$ & $1.55652$ & $21.9231$ & $1.68522$ \\
        ELG & $9.342 \cdot 10^{169}$ & $83.3254$ & $383.703$ & $0.219326$ \\ \hline
    \end{tabular}\vspace{0.15in}
    \caption{Best-fitting parameters for the differential galaxy number of the two DESI samples using the model function (\ref{eq:dNdz_fit}).
    }\vspace{-0.15in}
    \label{tab:dNdz_fit_parameters}
\end{table}

\begin{table}[t]
    \begin{minipage}[t]{.5\textwidth}
        \centering
        \begin{tabular}{c|c}
            \hline
            Parameter & Fiducial value \\ \hline
            $\omega_{\rm b}$ & $0.02237$ \\
            $\omega_{\rm cdm}$ & $0.1200$ \\
            $h$ & $0.6736$ \\
            $\mathrm{ln} \left( 10^{10} A_{\rm s} \right)$ & $3.044$ \\
            $\sigma_8$ & $0.82307$ \\
            $n_{\rm s}$ & $0.9649$ \\
            $w_0$ & $-1$ \\
            $w_a$ & $0$ \\ \hline
        \end{tabular}
    \end{minipage}
    \begin{minipage}[t]{.5\textwidth}
        \centering
        \begin{tabular}{c|c}
            \hline
            Parameter & Fiducial value \\ \hline
            $c_0 \, \left[ \mathrm{Mpc}^2 \right]$ & $-10$ \\
            $c_1 \, \left[ \mathrm{Mpc}^2 \right]$ & $20$ \\
            $c_2 \, \left[ \mathrm{Mpc}^2 \right]$ & $20$ \\
            $c_3 \, \left[ \mathrm{Mpc}^2 \right]$ & $20$ \\
            \hdashline
            $s_{0}$ & $0$ \\
            $s_{2}\, \left[ \mathrm{Mpc}^{-2} \right]$ & $0$ \\
            $s_{3}\, \left[ \mathrm{Mpc}^{-4} \right]$ & $0$ \\
            & \\ \hline
        \end{tabular}
    \end{minipage}
    \vspace{0.15in}
    \caption{Fiducial parameter values for the MCMC forecasts of the power spectrum wedges from a DESI-like experiment. The fiducial time-dependence of the biases are depicted in fig. \ref{fig:galaxy_number_rebinning}. The stochastic and the counter terms are assumed to be redshift independent, with fiducial values set to the inferred values from the {\sc Abacus} mocks.}
    \label{tab:1loop-fiducial}
    \vspace{-0.2in}
\end{table}

Since we consider 4 redshift bins for ELGs and 2 for BGs, with 12 nuisance parameters in each redshift (4 biases, 4 counter terms, 4 stochastic terms) and 7 free cosmological parameters, our runs should in principle feature a total of 79 free parameters. In order to have numerically tractable forecasts, like in {\sc Abacus} likelihood analysis, we set the $k^2$ stochastic contribution to zero ($s_1=0$) at all redshifts. Furthermore, we reduce the size of the parameter space by fixing the redshift dependence of some of the nuisance parameters and varying only overall factors for each of the two samples. More specifically, for each sample ${\rm x}$, we vary four parameters $({\tilde b}_1^{\rm x}, {\tilde b}_2^{\rm x}, {\tilde b}_{\mathcal{G}_2}^{\rm x}, {\tilde b}_{\Gamma_3}^{\rm x})$ such that the redshift-dependent biases read
\begin{align}
    b_1^{\rm x}(z) &= {\tilde b}_1^{\rm x} \times b_1^{\rm x}(z)~,   \label{eq:b1z} \\
    b_2^{\rm x}(z) &= \tilde{b}_2^{\rm x} \times \left( -0.704 - 0.208 z + 0.183 z^2 - 0.00771 z^3 \right)~,     \label{eq:b2z} \\
    b_{\cG_2}^{\rm x}(z) &= - \tilde{b}_{\cG_2}^{\rm x}  \times \frac{2}{7}\left[b_1^{\rm x}(z) - 1 \right]\label{eq:bG2z} \\
    b_{\Gamma_{3}}^{\rm x}(z) &= \tilde{b}_{\Gamma_{3}}^{\rm x} \times \frac{23}{42} \left[b_1^{\rm x}(z) - 1 \right]~. \label{eq:bG3z}
\end{align}
    
The redshift dependence of the linear bias $b_1^{\rm x}(z)$, shown in the middle panel of figure \ref{fig:galaxy_number_rebinning}, is based on ref. \cite{DESI:2016fyo}. The redshift dependence of $b_2$ is set according to refs. \cite{Yankelevich:2018uaz,DiDio:2018unb}, while for $b_{\cG_2}$ and $b_{\Gamma_3}$ we follow the co-evolution prediction. We further assume that the redshift evolution of the stochastic contributions $s_n$ is the same as that of the Poisson shot noise $1/{\bar n}(z)$, and only vary three overall factors $s_n$ for each sample.
Finally, In the absence of a well-motivated assumption for the redshift evolution of counter terms, we should in principle vary four independent parameters $c_n$ for each redshift bin and each galaxy sample. However, in our DESI forecasts, for computational tractability, we assume that counter terms are redshift independent and only vary four parameters $c_n$ for each of the two galaxy samples. These assumptions reduce the total number of free parameters to $2 \times 11 + 7 = 29$. For both counter-terms and shot noise corrections, we consider the same fiducial values for the two samples, chosen to be of the same order as the ones inferred from {\sc Abacus}. The fiducial values are summarised in table \ref{tab:1loop-fiducial}. 

\begin{figure}[t]
    \centering
    \includegraphics[width=0.7\textwidth]{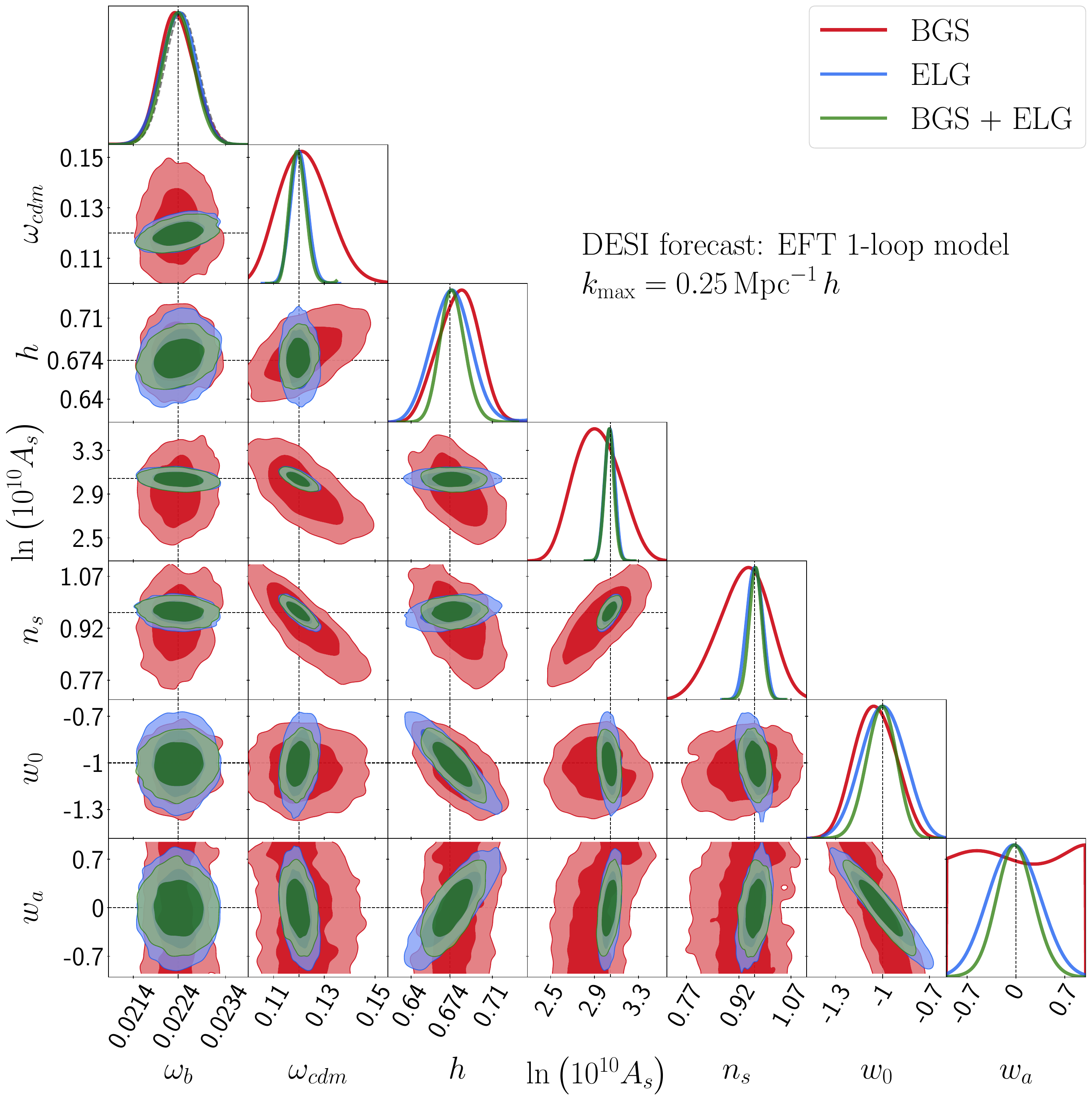}
    \caption{Forecasted marginalized posterior distributions of cosmological parameters from BGS (red) and ELG (blue) samples and their combination (green) for a DESI-like survey. The synthetic power spectrum wedges are generated using the one-loop EFT model of the fiducial model at the corresponding redshifts for the two samples. The small-scale cutoff is set to $k_{\mathrm{max}} = 0.25 \, \mathrm{Mpc}^{-1}\, h$. }
    \label{fig:1loop-bgs_elg_full_cosmo}
\end{figure}

\subsection{Results: constraints on the $w$CDM model}\label{sec:DESI_res}

Figure \ref{fig:1loop-bgs_elg_full_cosmo} shows a comparison of the marginalized posterior distributions of cosmological parameters derived from our mock DESI-like survey using power spectrum wedges. The small-scale cutoff is set to $k_{\rm max} = 0.25 \ h/{\rm Mpc}$. The constraints from the BGS and ELG samples and their combination are shown in red, blue, and green, respectively. For $\{\omega_{\rm b}, h, w_0\}$, the two samples provide comparable parameter uncertainties. However, the ELG sample, which spans higher redshifts, provides significantly tighter constraints on $\{\omega_{\rm cdm}, \ln( 10^{10} A_{\rm s}), n_{\rm s}, w_{a}\}$. The combination of the two samples considerably improves the constraints on $\{h, w_0, w_a\}$, by about 35\% for the first two parameters and about 30\% for $w_a$. The slightly rotated degeneracy direction in the $w_0-w_a$ plane plays an important role in obtaining improved constraints on the dark energy equation of state.

\begin{figure}[htbp!]
    \centering
    \includegraphics[width=0.7\textwidth]{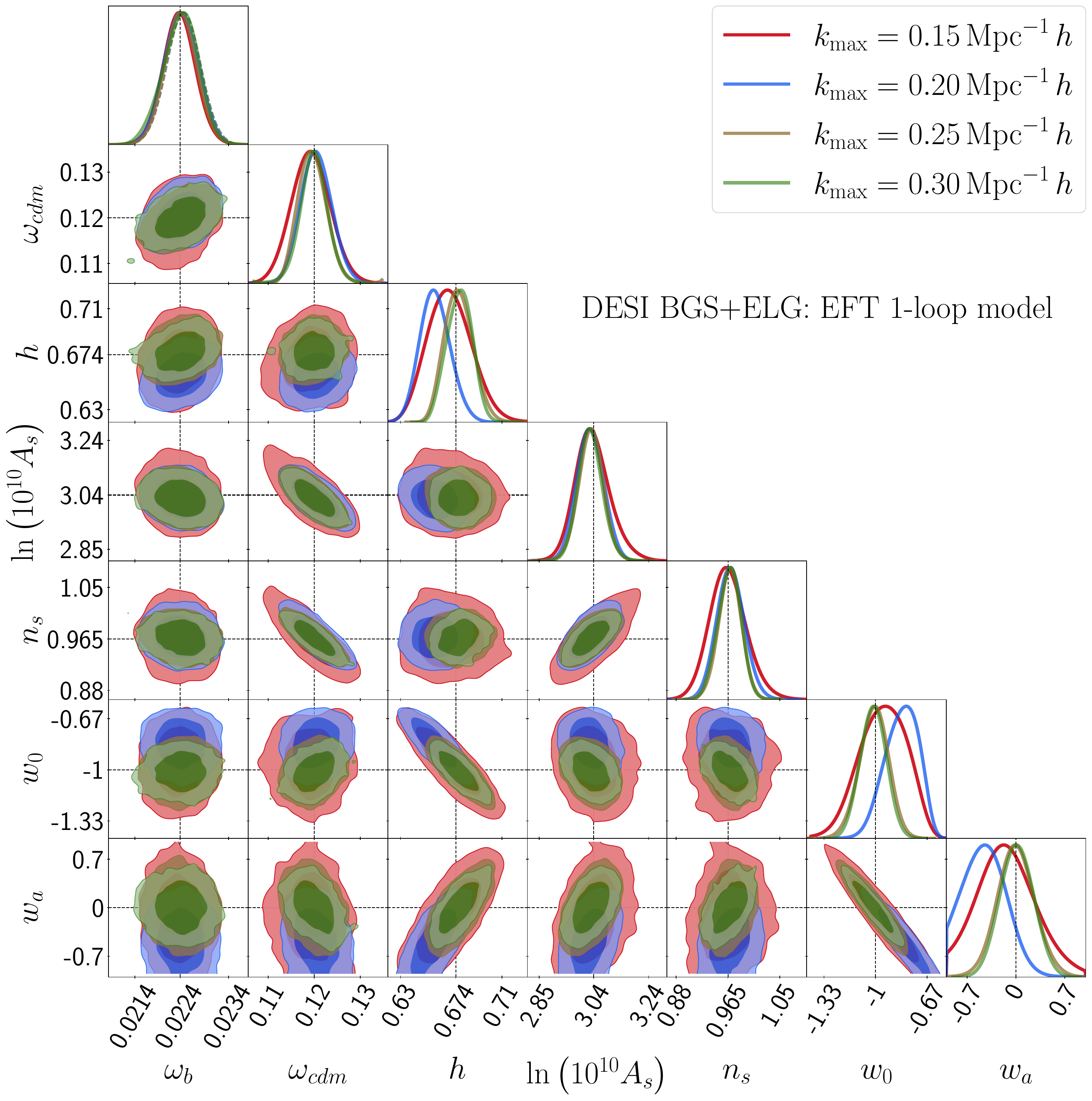}
    \caption{Forecastsed marginalized posterior distributions from the full DESI-like sample for different values of $k_{\mathrm{max}}$.} 
    \label{fig:1loop-kmax_cosmo}
\end{figure}

Figure \ref{fig:1loop-kmax_cosmo} illustrates the dependence of the constraints on the choice of small-scale cutoff. The different colors correspond to four values of $k_{\rm max}[h/{\rm Mpc}]=\{0.15,0.2,0.25,0.3\}$. We also show in figure~\ref{fig:desi-1loop-full} of appendix \ref{app:nuisance} the full contours on all cosmological and nuisance parameters in the cases $k_{\rm max}[h/{\rm Mpc}]=\{0.15,0.25\}$.

The constraints on $\omega_{\rm b}$ are fully determined by the BBN prior, thus, we do not see any improvement when increasing $k_{\rm max}$, while we do for all other parameters. Imposing a highly conservative cutoff of $k_{\rm max}=0.15 \ h/{\rm Mpc}$ results in considerably weaker constraints. Indeed, discarding the modes with $k>0.15 \, h/{\rm Mpc}$ leaves the nuisance parameters poorly constrained, as can be checked in figure \ref{fig:desi-1loop-full}. These degrees of freedom are then redundant and the predictive power of the model is reduced. Beyond $k_{\rm max} = 0.25 \ h/{\rm Mpc}$, the constraints on $w_0$ and $w_a$ show 10\% improvements, while for other parameters, the improvement is negligible. 

\begin{table}[htbp!]
    \centering
    \begin{tabular}{c|ccc}
        \hline
       Parameter & BGS & ELG & BGS $+$ ELG  \\
       & (z-dep fixed) & (z-dep fixed) & (z-dep fixed)  \\ \hline
        $\omega_{\rm b}$ & $3.7\times 10^{-4}$ & $3.7\times 10^{-4}$ & $3.6\times 10^{-4}$  \\
        $\omega_{\rm cdm}$ & $0.0098$ & $0.0033$ & $0.0030$   \\
        $h$ & $0.017$ & $0.017$ & $0.011$  \\
        $\mathrm{ln} \left( 10^{10} A_{\rm s} \right)$ & $0.21$ & $0.047$ & $0.046$  \\
        $n_{\rm s}$ & $0.072$ & $0.022$ & $0.019$  \\
        $w_0$ & $0.14$ & $0.14$ & $0.094$ \\
        $w_a$ & -- & $0.36$ & $0.27$  \\ \hline
        $\sigma_8$ & $0.042$ & $0.015$ & $0.013$  \\ \hline
    \end{tabular}\vspace{0.15in}
    \caption{Forecasted 1$\sigma$ errors on independently varied cosmological parameters, as well as $\sigma_8$ as a derived parameter. We show the results for the BGS and ELG samples individually and combined, assuming a fixed redshift-dependence of nuisance parameters is fixed. Here, the small-scale cutoff is set to $k_{\mathrm{max}} = 0.25 \,h/\mathrm{Mpc}^{-1}$. Since the BGS sample alone can not constrain $w_a$ within the prior range, the corresponding cell is left empty.}\vspace{-0.2in}
    \label{tab:forecast_sigmas}
\end{table}

The the full triangle plot in fig. \ref{fig:desi-1loop-full} of appendix \ref{app:nuisance} shows the level of correlation between cosmological and nuisance parameters.
Overall, compared to the $\Lambda$CDM fits to {\sc Abacus} mocks, the additional cosmological parameters corresponding to the DE equation of state do not exhibit any degeneracy with nuisance parameters. In addition to linear bias, which is anti-correlated with $A_{\rm s}$, the stochastic parameter $s_0$ is the most correlated nuisance parameter with cosmological parameters, particularly with $\omega_{\rm b},\omega_{\rm cdm}, A_{\rm s}$, and $n_{\rm s}$.

Table \ref{tab:forecast_sigmas} presents a final summary of our results on the sensitivity of a DESI-like survey to cosmological parameters, using a one-loop modelling of power spectrum wedges. We show here the 1$\sigma$ error bar on each parameter using each of the two galaxy samples or their combination, always with a cutoff at $k_{\rm max} = 0.25 \ h/{\rm Mpc}$. As described earlier, the ELG sample provides tighter constraints on all parameters. Combining the two samples improves the constraints, most notably on the Hubble parameter and dark energy equation of state.

\section{Conclusion}\label{sec:conc}

The cosmic large-scale structure contains a rich trove of information about the initial conditions of the universe, its constituents, and the laws of gravity determining its evolution. The upcoming data from stage-IV galaxy surveys has the potential to substantially improve our understanding of the Universe, thanks to the the enhanced precision and volume of the data compared to current surveys. Exploiting these rich datasets to their potential will require an accurate theoretical modeling of the observables, as well as some fast and robust numerical implementation to compute the theoretical model and perform cosmological inference. In recent years, the modeling of LSS summary statistics and the numerical tools to analyze the measurements, in particular at the level of the power spectrum, have made considerable progress. Several well-tested codes have been applied to existing data and released publicly. There are ongoing efforts in the community to enhance the precision, robustness, and efficiency of these tools. 

In this paper, we introduced {\sc CLASS-OneLoop}, a new numerical tool fully integrated in the widely used {\sc CLASS} Boltzmann code, to compute the one-loop power spectrum of biased tracers in redshift space. We validated our implementation by comparing the results of the FFTLog approach versus a direct numerical integration of loop integrals. Furthermore, we performed an explicit comparison of {\sc CLASS-OneLoop} against the publicly available code {\sc CLASS-pt}. We showed that when 
 the same algorithm is plugged into the two codes to split the linear matter power spectrum into broadband and BAO terms (in order to perform the IR resummation), these two codes agree to better than 0.3\% up to $k=0.3\,h/{\rm Mpc}$.

To illustrate the potential of {\sc CLASS-OneLoop}, we used the new pipeline together with the package {\sc MontePython} to infer cosmological parameters from the measurement of power spectrum wedges on a $\Lambda$CDM subset of the {\sc AbacusSummit} simulation suite. We found that one-loop EFT predictions are capable of nearly perfectly recovering the fiducial cosmology of the simulations. For comparison, we performed a similar exercise with a simple phenomenological model used in the literature for sensitivity forecasts, and found highly biased constraints with deceptively small error bars. This shows that the one-loop model is robust and accurate, while the simplified model should not be used to fit real data.

Next, we used our inference pipeline to perform Monte Carlo forecasts and determined the constraining power of  stage-IV spectroscopic galaxy surveys. We estimated the sensitivity of such a survey to a cosmological model featuring dynamical dark energy, assuming the simple $w_0-w_a$ parameterization \cite{Chevallier:2000qy}. Adopting DESI-like sensitivity settings and imposing a BBN prior on $\omega_{\rm b}$, we found that a DESI-like experiment can set tight constraints on dark energy and $\Lambda$CDM parameters, despite of the large number of free parameters in the one-loop EFT model. Even with 22 model parameters (describing bias, EFT counter-term and stochastic uncertainties in each sample), our MCMC chains converge towards definite predictions for each cosmological and model parameter, with limited correlations between the two types of parameters.

The technical implementation of the one-loop formalism into {\sc CLASS} and a more detailed description of numerical algorithms and performance will be the topic of a work in preparation~\cite{release}. {\sc CLASS-OneLoop} will be publicly released (as part of the main {\sc CLASS} distribution) together with that publication.

\subsection*{Acknowledgements}
We thank Chang-hoon Hahn for providing us the {\sc Abacus} sub-boxes. We also thank Zvonomir Vlah, Marko Simonovic, and Guido D'Amico for helpful discussions. The likelihood analysis of {\sc  Abacus} simulations and the forecasts were performed with computing resources granted by RWTH Aachen University under project rwth1389. The measurements of the power spectrum wedges were performed on the {\it Rusty} cluster at the Flatiron Institute.

\appendix

\section{Simplified phenomenological model for the nonlinear power spectrum}
\label{app:SPM}

As a comparison point for the base one-loop model described in section~\ref{subsec:EFT}, we consider a simple phenomenological model (SPM) for the nonlinear power spectrum of biased tracers. This model has been previously used in the literature in performing Fisher or MCMC forecasts (e.g., \cite{Euclid:2019clj, Euclid:2023pxu}), and while it is clearly too simplistic, as we will discuss when fitting the models to N-body simulations, it serves the purpose of highlighting the importance of various ingredients of the one-loop model. 

The SPM consists of a clustering and a constant shot-noise components. The clustering contribution assumes a linear biasing relation between the DM and the tracer, accounts for linear redshift-space distortions by including the Kaiser term \cite{Kaiser:1987qv}, and models the FoG effect with a Lorentzian suppression of power  \cite{Ballinger:1996cd}. Furthermore, it models the damping of the BAOs due to large displacement fields using velocity dispersion as the damping exponent. The shot-noise component, $P_s$, is described as a free parameter to account for possible sub- or super-Poissonian corrections. Putting all this together, the redshift-space galaxy power spectrum of the SPM reads
\be\label{eq:GC:pk-ext}
P^s_g(k,\mu) = \left[\frac{(b+f\mu^2)^2 }{1+f^2 k^2\mu^2\sigma_{\rm p}^2}\right]\left[P_{\rm nw}(k)+e^{-\sigma_{\rm v}^2(\mu)k^2}P_{\rm w}(k) \right] + P_\text{s} \, , 
\ee
where, like in previous sections, we have omitted the explicit redshift dependence. The model also includes the AP effect as described in section~\ref{subsub:AP}.
In analyzing {\sc Abacus} data, we vary independently $\sigma_{\rm p}$, which characterizes the FoG suppression. As in ref. \cite{Euclid:2023pxu}, we assume the dependence of $\sigma_{\rm v}$ on LoS direction to be given by
\begin{equation}
\sigma^2_{\rm v}(\mu) =  \left\{1 + f(f+2)\mu^2 \right\} \tilde \sigma_{\rm v}^2~,
\label{eq:sigmav}
\end{equation}
which matches the expression of $\Sigma_{\rm s}^2(\mu,z)$ in eq.~\eqref{eq:rsd_damp} if we identify $\tilde \sigma_{\rm v}=\Sigma$ and set $\delta \Sigma^2 = 0$. In fitting the {\sc ABACUS} mocks, we found that if varying $\tilde \sigma_{\rm v}$ as a free parameter, the inferred values of several cosmological parameters, most notably $h$, drift from their expected values. Therefore, in the presented analysis, we fix it to be $\tilde \sigma_{\rm v}=\Sigma$.
 
\section{Anisotropic \texttt{FFTLog} Kernels for the LoS moments} 
\label{app:FFTLog}

The following ansatze and end results can also be found in \cite{Chudaykin:2020aoj}.\footnote{Whe have confirmed with the authors of \cite{Chudaykin:2020aoj} a typo within their B2 term.} To solve higher LoS moments, we have to build recursive relationships on top of the kernel $I(\nu_1,\nu_2)$ defined as \cite{Simonovic:2017mhp}
\begin{equation}
    k^{3-2(\nu_1+\nu_2)}I(\nu_1,\nu_2) = \int \dd^3\vec{q}\quad \frac{1}{q^{2\nu_1}|\vec{k} -\vec{q}|^{2\nu_2}}~,
\end{equation}
with
\begin{equation}
    I(\nu_1,\nu_2)=\frac{1}{8\pi^{3/2}}\frac{\Gamma(\frac{3}{2}-\nu_1)\Gamma(\frac{3}{2}-\nu_2)\Gamma((\nu_1+\nu_2)\frac{3}{2})}{\Gamma(\nu_1)\Gamma(\nu_2)\Gamma(3 - (\nu_1+\nu_2))}~.
\end{equation}   

To obtain terms like, for example, $A_1(\nu_1,\nu_2)$, we must prove that such terms are aligned with the wave vector $\vec{k}$. This follows from
 \begin{equation}
        \int \dd^3\vec{q}\quad \frac{q_i}{q^{2\nu_1}|\vec{k} -\vec{q}|^{2\nu_2}}
        = \int \dd^3\vec{q}\quad \frac{q_i}{q^{2\nu_1}k^{2\nu_2}} \sum^{\infty}_{n=0} \left( -\frac{q^2}{k^2} + \frac{2\vec{k}\cdot\vec{q}}{k^2}\right)^{n\nu_2}
\end{equation}
Due to rotational symmetry of the integrand along $\vec{k}$, it becomes evident that the integral aligns itself with respect to the external wave vector, leading to the following ansatz,
\begin{equation}
    \int \dd^3\vec{q}\quad \frac{q_i}{q^{2\nu_1}|\vec{k} -\vec{q}|^{2\nu_2}} = k^{3-2(\nu_1+\nu_2)} k_i A_1(\nu_1,\nu_2)~.
\end{equation}
By contracting both sides with $k^i$, we obtain
\begin{gather*}
    k^{3-2(\nu_1+\nu_2)}  A_1(\nu_1,\nu_2) = \int \dd^3\vec{q}\quad \frac{\vec{q} \cdot \vec{k}}{k^2}\frac{1}{q^{2\nu_1}|\vec{k} -\vec{q}|^{2\nu_2}} \\
    \int \dd^3\vec{q}\quad \frac{1}{q^{2\nu_1}|\vec{k} -\vec{q}|^{2\nu_2}} \frac{1}{2k^2}\left( -|\vec{k} -\vec{q}|^2 + k^2 +q^2\right)\\
    = \frac{1}{2}(I(\nu_1,\nu_2)+I(\nu_1-1,\nu_2)-I(\nu_1,\nu_2-1)) k^{3-2(\nu_1+\nu_2)}\\
    \Rightarrow A_1(\nu_1,\nu_2) = \frac{1}{2}(I(\nu_1,\nu_2)+I(\nu_1-1,\nu_2)-I(\nu_1,\nu_2-1))~.
\end{gather*}
Similar considerations lead to the following expression of the additional anisotropic \texttt{FFTLog} kernels:

\begin{align}
     A_2(\nu_1,\nu_2)&= \frac{1}{8} (2I(\nu_1-1,\nu_2-1)+2I(\nu_1-1,\nu_2)+2I(\nu_1,\nu_2-1) \nonumber\\
     &-I(\nu_1-2,\nu_2)-I(\nu_1,\nu_2-2)-I(\nu_1,\nu_2)) \\
     B_2(\nu_1,\nu_2)&=\frac{1}{8} (-6I(\nu_1-1,\nu_2-1)+2I(\nu_1-1,\nu_2)-6I(\nu_1,\nu_2-1) \nonumber\\
     &+3I(\nu_1-2,\nu_2)+3I(\nu_1,\nu_2-2)+3I(\nu_1,\nu_2))
\end{align}
\begin{align}
     A_3(\nu_1,\nu_2)&=-\frac{3}{16} (I(\nu_1-3,\nu_2)-3 I(\nu_1-2,\nu_2-1)-I(\nu_1-2,\nu_2)+3I(\nu_1-1,\nu_2-2) \nonumber\\
     &-2 I(\nu_1-1,\nu_2-1)-I(\nu_1-1,\nu_2)-I(\nu_1,\nu_2-3)+3I(\nu_1,\nu_2-2) \nonumber \\
     &-3 I(\nu_1,\nu_2-1)+I(\nu_1,\nu_2)) \\
     B_3(\nu_1,\nu_2)&=\frac{1}{16} (5I(\nu_1-3,\nu_2)-15I(\nu_1-2,\nu_2-1)+3 I(\nu_1-2,\nu_2)+15I(\nu_1-1,\nu_2-2) \nonumber \\
     &-18I(\nu_1-1,\nu_2-1)+3 I(\nu_1-1,\nu_2)-5I(\nu_1,\nu_2-3)+15I(\nu_1,\nu_2-2) \nonumber \\
     &-15I(\nu_1,\nu_2-1)+5I(\nu_1,\nu_2)) 
\end{align}
\begin{align}
     A_4(\nu_1,\nu_2)&=\frac{3}{128} (I(\nu_1-4,\nu_2)-4 I(\nu_1-3,\nu_2-1)-4 I(\nu_1-3,\nu_2)+6I(\nu_1-2,\nu_2-2) \nonumber\\
     &+4 I(\nu_1-2,\nu_2-1)+6 I(\nu_1-2,\nu_2)-4 I(\nu_1-1,\nu_2-3)+4 I(\nu_1-1,\nu_2-2)\nonumber\\
     &+4I(\nu_1-1,\nu_2-1)-4 I(\nu_1-1,\nu_2)+I(\nu_1,\nu_2-4)-4 I(\nu_1,\nu_2-3)\nonumber\\
     &+6 I(\nu_1,\nu_2-2)-4 I(\nu_1,\nu_2-1)+I(\nu_1,\nu_2)) \\
     B_4(\nu_1,\nu_2)&=-\frac{3}{64} (5 I(\nu_1-4,\nu_2)-20 I(\nu_1-3,\nu_2-1)-4 I(\nu_1-3,\nu_2)+30I(\nu_1-2,\nu_2-2) \nonumber\\
     &-12I(\nu_1-2,\nu_2-1)-2I(\nu_1-2,\nu_2)-20 I(\nu_1-1,\nu_2-3)+36 I(\nu_1-1,\nu_2-2) \nonumber\\
     &-12I(\nu_1-1,\nu_2-1)-4 I(\nu_1-1,\nu_2)+5I(\nu_1,\nu_2-4)-20 I(\nu_1,\nu_2-3) \nonumber\\
     &+30 I(\nu_1,\nu_2-2)-20I(\nu_1,\nu_2-1)+5 I(\nu_1,\nu_2)) \\
     C_4(\nu_1,\nu_2)&=\frac{1}{128} (35 I(\nu_1-4,\nu_2)-140 I(\nu_1-3,\nu_2-1)+20 I(\nu_1-3,\nu_2)+210I(\nu_1-2,\nu_2-2) \nonumber \\
     &-180 I(\nu_1-2,\nu_2-1)+18 I(\nu_1-2,\nu_2)-140I(\nu_1-1,\nu_2-3)+300 I(\nu_1-1,\nu_2-2)\nonumber\\
     &-180 I(\nu_1-1,\nu_2-1)+20I(\nu_1-1,\nu_2)+35 I(\nu_1,\nu_2-4)-140 I(\nu_1,\nu_2-3) \nonumber\\
     &+210 I(\nu_1,\nu_2-2)-140 I(\nu_1,\nu_2-1)+35 I(\nu_1,\nu_2))
\end{align}

\section{Notation for the contributions to one-loop contributions}
\label{app:loops}

The numerical implementation of one-loop contributions to the redshift-space halo power spectrum follows the notation described below. The loops corresponding to correlations of two (density or velocity) fields are referred to as $\cI$, while those with three and four fields are labeled $\cJ$ and $\cN$, respectively. We use subscripts to refer to the number of velocity fields. We have two types of terms, those that involve nonlinear biases, i.e. bias loops, and those that are due to non-linearities in DM density and velocity fields. For matter loops, we use superscripts to refer to the perturbative orders of density/velocity fields, while for the bias terms, we use superscripts with the name of the nonlinear bias operators. With these conventions, the contributions to velocity moments listed in Eqs. \eqref{eq:pnn_loops} read in terms of individual loop integrals as:
\begingroup
\allowdisplaybreaks
\begin{align}
P_{00}^{\rm loop}(k) &= 2b_1^2 \cI_{00}^{22}(k) + 6 b_1^2  P_0(k) \cI_{00}^{13}(k) + 2b_1 b_2 \cI_{00}^{\delta^2}(k)+ 4 b_1 b_{\cG_2} \cI_{00}^{\cG_2}(k) \notag \\
&+ \frac{1}{2} b_2^2 \cI_{00}^{\delta^2 \delta^2} + 2b_{\cG_2}^2\cI_{00}^{\cG_2 \cG_2} +  2 b_2 b_{\cG_2} \cI_{00}^{\delta^2 \cG_2}(k) + 8 b_1 (b_{\cG_2} + \frac{2}{5}b_{\Gamma_3})  P_0(k) \cF_{00}^{\cG_2}(k)~, \\
P_{01}^{\rm loop}(k,\mu) &= \left(-if \frac{\mu}{k} \right) \Bigg\{ 2b_1 \cI_{01}^{22}(k) + 3 b_1 P_0(k) \Big[\cI_{01}^{13}(k) + \cI_{01}^{31}(k)\Big] + b_2 \cI_{01}^{\delta^2}(k)+ 2 b_{\cG_2} \cI_{01}^{\cG_2}(k) \Bigg . \notag  \\ 
&+ \Bigg.  4 (b_{\cG_2} + \frac{2}{5}b_{\Gamma_3}) P_0(k) \cF_{01}^{\cG_2}(k) \Bigg\} 
-2 i f  \Bigg\{ b_1^2 P_0(k) \Big[\cJ_{01}^{121}(k,\mu) +  \cJ_{01}^{112}(k,\mu)\Big]  \Bigg.  \notag \\
& \Bigg . \quad \quad +   b_1^2\cJ_{01}^{211}(k,\mu) +  \frac{1}{2} b_1 b_2\cJ_{01}^{\delta_2}(k,\mu) + b_1 b_{\cG_2}\cJ_{01}^{\cG_2}(k,\mu)\Bigg\}~, \\
P_{02}^{\rm loop}(k,\mu) &=   -f^2 \Bigg\{ 4b_1 P_0(k) \cJ_{02}^{121}(k,\mu) + 2b_1  \cJ_{02}^{211}(k)  +  b_2 \cJ_{02}^{\delta^2}(k,\mu) + 2 b_{\cG_2} \cJ_{02}^{\cG_2}(k,\mu) - b_1^2 P_0(k) \sigma_v^2 \Bigg\}~, \\
P_{11}^{\rm loop}(k,\mu) &=  f^2 \left(\frac{\mu}{k}\right)^2 \Bigg\{ 2 \cI_{11}^{22}(k,\mu) + 6 P_0(k) \cI_{11}^{13}(k,\mu)\Bigg\} +  f^2 b_1^2\cN_{11}^{1111}(k,\mu) \notag \\
&+ 4 f^2 \left(\frac{\mu}{k}\right) \Bigg\{b_1 P_0(k)\Big[\cJ_{11}^{121}(k,\mu) + \cJ_{11}^{112}(k,\mu)\Big] + b_1 \cJ_{11}^{211}(k,\mu) \Bigg\}, \\
P_{03}^{\rm loop}(k,\mu) &= -3 if^3 \left(\frac{\mu}{k}\right) b_1 P_0(k) \sigma_v^2~, \\
P_{12}^{\rm loop}(k,\mu) &= -2 if^3 \left(\frac{\mu}{k}\right)\Bigg\{ \cJ_{12}^{211}(k,\mu) + 2 P_0(k) \cJ_{12}^{121}(k,\mu) \Bigg\} \notag \\
&+ i f^3 \left(\frac{\mu}{k}\right) b_1 P_0(k)  \sigma_v^2 -  2 i f^3 b_1 \cN_{12}^{1111}(k,\mu)~, \\
P_{13}^{\rm loop}(k,\mu) &= 3 f^4 \left(\frac{\mu}{k}\right)^2 P_0(k) \sigma_v^2~, \\
P_{22}^{\rm loop}(k,\mu) &= 2 f^4 \cN_{22}^{1111}(k,\mu)~.
\end{align}
\endgroup
There are three tree-level contributions,
\begin{align}
P_{00}^{\rm tree}(k,\mu) &= b_1^2 P_0(k)~,\\
P_{01}^{\rm tree}(k,\mu) &= \left(-if \frac{\mu}{k} \right) b_1 P_0(k)~, \\
P_{11}^{\rm tree}(k,\mu) &= f^2 \left(\frac{\mu}{k}\right)^2 P_0(k). 
\end{align}

\section{Full cosmological and nuisance parameter contours}
\label{app:nuisance}

\begin{figure}[h]
    \centering
    \includegraphics[width=0.9 \textwidth]{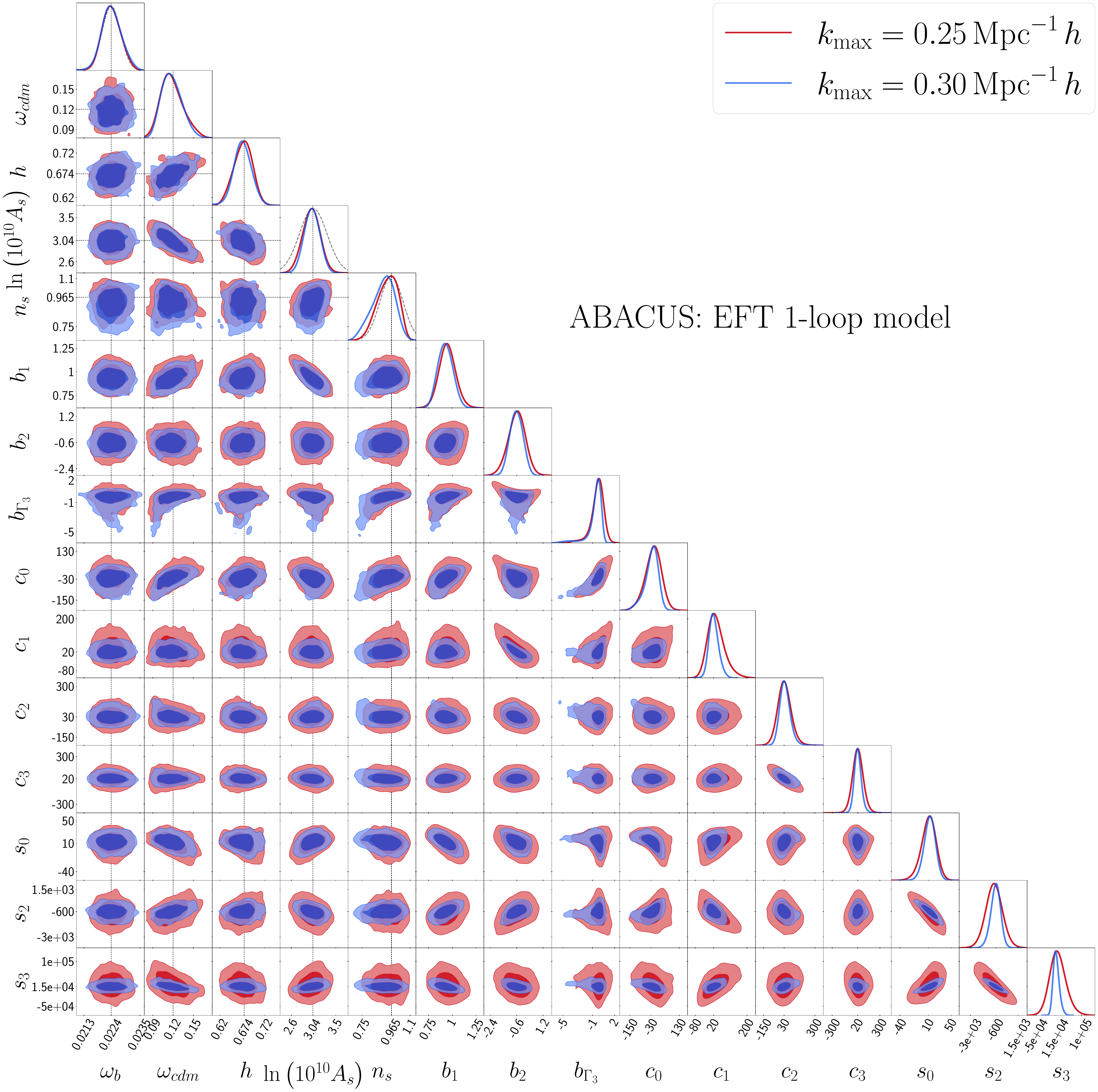}
    \caption{Marginalized posterior distributions for all parameters of the one-loop EFT model inferred from the power spectrum wedges extracted from the {\sc Abacus} simulation suite at redshift of $z = 0.5$. The small-scale cutoff is set to $k_{\rm max} = 0.25 \, h \, \mathrm{Mpc}^{-1}$ for the red and to $k_{\rm max} = 0.3 \, h \, \mathrm{Mpc}^{-1}$ for the blue contours.}
    \label{fig:abacus-1loop-full}
\end{figure}

\begin{figure}[t]
    \centering
    \includegraphics[width=\textwidth]{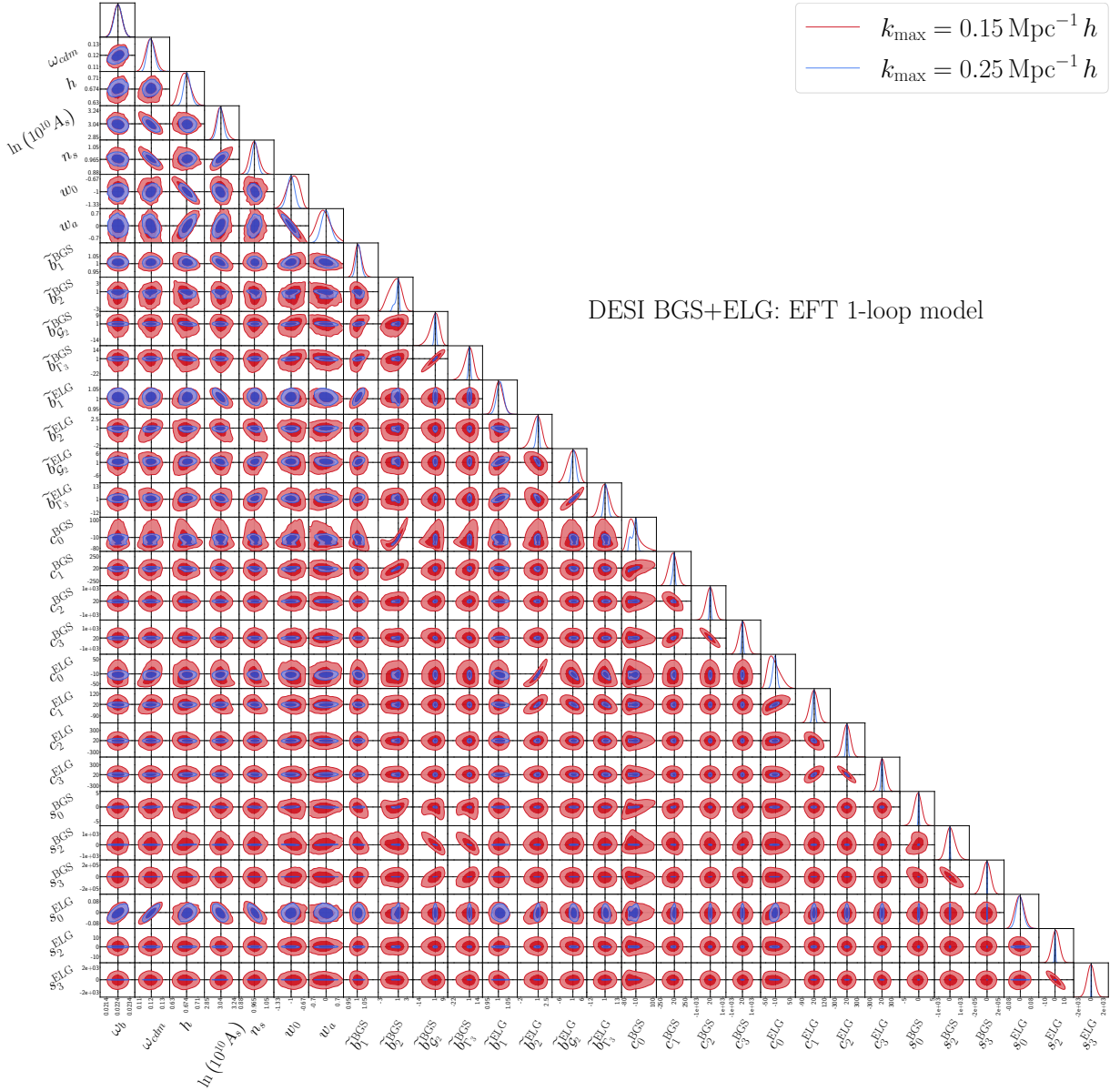}
    \caption{Forecasted marginalized posterior distributions of all model parameters from the combination of BGS and ELG samples for a DESI-like survey. The synthetic power spectrum wedges are generated using the one-loop EFT model of the fiducial model at the corresponding redshifts for the two samples. The small-scale cutoff is set to $k_{\mathrm{max}} = 0.15\, \mathrm{Mpc}^{-1}\, h$ for the red and to $k_{\mathrm{max}} = 0.25\, \mathrm{Mpc}^{-1}\, h$ for the blue contours.}
    \label{fig:desi-1loop-full}
\end{figure}

To illustrate the degeneracies among nuisance and cosmological parameters, we present in this section the full triangle plots of 1d and 2d marginalized parameter posteriors for fits to {\sc Abacus} mocks in figure \ref{fig:abacus-1loop-full} and DESI-like MCMC forecasts in figure \ref{fig:desi-1loop-full}. \\

\newpage
\bibliographystyle{utphys}
\bibliography{Class1loop_v2}

\end{document}